\documentclass[prd,superscriptaddress,preprintnumbers,amsmath,nofootinbib,preprint]{revtex4}
\usepackage{amsmath, amssymb, graphics,epsfig}

\usepackage{slashed}

\usepackage{hyperref}

\def\bfnabla{\mbox{\boldmath $\nabla$}}

\def\bfsigma{\mbox{\boldmath $\sigma$}}
\def\lQ{\Lambda_{\rm QCD}}
\newcommand{\nn}{\nonumber}
\newcommand{\be}{\begin{equation}}
\newcommand{\ee}{\end{equation}}
\newcommand{\bea}{\begin{eqnarray}}
\newcommand{\eea}{\end{eqnarray}}
\def\al{\alpha}

\def\siml{{\
    \lower-1.2pt\vbox{\hbox{\rlap{$<$}\lower6pt\vbox{\hbox{$\sim$}}}}\ }}

\def\dsl{\,\raise.15ex\hbox{/}\mkern-13.5mu D}

\def\bfnabla{\mbox{\boldmath $\nabla$}}
\def\bfsigma{\mbox{\boldmath $\sigma$}}

\def\lQ{\Lambda_{\rm QCD}}

\newcommand{\eps}{\epsilon}

\newcommand{\cf}{C_F}

\newcommand{\Appendix}[1]%
    {%
     \section{#1}%
      }

\begin{document}

\title{Spectator effects in the HQET renormalization group improved Lagrangian at $\mathcal{O}(1/m^3)$ with leading logarithmic accuracy: Spin-dependent case}
\author{Daniel Moreno}
\email{dmoreno@ifae.es}
\address{Grup de F\'\i sica Te\`orica, Dept. F\'\i sica and IFAE-BIST, Universitat Aut\`onoma de Barcelona,\\ 
E-08193 Bellaterra (Barcelona), Spain}
\date{\today}

\begin{abstract}
The present paper incorporates the effects induced by massless spectator quarks in the renormalization group improved Wilson coefficients associated to the  
${\cal O}(1/m^3)$ spin-dependent heavy quark effective theory Lagrangian operators. The computation is carried out in Coulomb gauge with 
leading logarithmic precision. The result completes the Lagrangian to $\mathcal{O}(1/m^3)$ with that accuracy. 
\end{abstract}
\maketitle

\section{Introduction}

The heavy quark effective theory (HQET) \cite{HQET} is a powerful tool used to describe hadrons containing a heavy quark. The HQET Lagrangian is also one 
of the building blocks of the nonrelativistic QCD (NRQCD) Lagrangian ~\cite{NRQCD,Bodwin:1994jh}, which aims to describe bound states of 
two heavy quarks, heavy quarkonium for short. Concurrently, the Wilson coefficients of the NRQCD Lagrangian enter into the Wilson coefficients 
(interaction potentials) of the potential NRQCD (pNQRCD) Lagrangian \cite{Pineda:1997bj,Brambilla:1999xf} as a consequence of the matching 
between the two effective theories. The later, is a theory optimised for the description of heavy quarkonium near 
threshold (for reviews see Refs. \cite{Brambilla:2004jw,Pineda:2011dg}).

Therefore, the Wilson coefficients computed in this paper could have many applications in heavy quark and heavy quarkonium physics. In particular, 
they are necessary ingredients to obtain the pNRQCD Lagrangian with next-to-next-to-next-to-next-to-leading order (NNNNLO) and 
with next-to-next-to-next-to-next-to-leading logarithmic (NNNNLL) accuracy, which is the necessary precision to determine the $\mathcal{O}(m\alpha^6)$ 
and the $\mathcal{O}(m\alpha^7\ln\alpha + m\alpha^8\ln^2\alpha + \ldots)$ heavy quarkonium spectrum.
As commented in Ref. \cite{lmp}, the Wilson coefficients associated to the $1/m^3$ spin-independent operators are also necessary to obtain 
the heavy quarkonium spectrum with NNNLL accuracy, and the production and annihilation of heavy quarkonium with NNLL 
precision \cite{inpreparation}, but this is not the case for the spin-dependent ones \cite{rgsddelta}. The Wilson coefficients computed in 
this paper also have applications in QED bound states like in muonic hydrogen. The computation presented here, also provides a cross-check that 
the physical combinations found in Ref. \cite{lmp} are gauge independent, and 
also new physical quantities involving light fermion operators are found. It also provides an additional cross-check of some of the reparametrization 
invariance relations given in Ref. \cite{Manohar:1997qy} and gives a solution in a more standard basis settled on by the same reference.

At present, the operator structure of the HQET Lagrangian, and the tree-level values of their Wilson coefficients, is known to ${\cal O}(1/m^3)$ in 
the case without spectator quarks \cite{Manohar:1997qy}. The inclusion of spectator quarks has been considered in Ref. \cite{Balzereit:1998am}.
The Wilson coefficients with leading logarithmic (LL) accuracy were computed in Refs. \cite{Finkemeier:1996uu,Blok:1996iz,Bauer:1997gs}
to ${\cal O}(1/m^2)$ and at next-to-leading order (NLO) in Ref. \cite{Manohar:1997qy} to ${\cal O}(1/m^2)$ (without heavy-light operators). The 
LL running to ${\cal O}(1/m^3)$ without the inclusion of spectator quarks was considered in Refs. \cite{Balzereit:1998jb,Balzereit:1998vh}, which turned out 
to have internal discrepancies between their explicit single log results and their own anomalous dimension matrix. The computation was reconsidered in 
Refs. \cite{chromopolarizabilities,lmp}, where the results were corrected. 

The inclusion of heavy-light operators to $\mathcal{O}(1/m^3)$ was considered 
in Ref. \cite{Balzereit:1998am}, but only single logarithmic results for the Wilson coefficients were provided. The inclusion of spectator quarks and the 
obtention of the full ressumed leading logarithmic (LL) expresions for the spin-independent case were obtained in Ref. \cite{chromopolarizabilities}. 
At the level of the single logs, these two references are in disagreement. This work is a follow-up to 
Refs. \cite{chromopolarizabilities,lmp} and, for this reason, is structured very similarly. The effect induced by massless
spectator quarks to the running of the $\mathcal{O}(1/m^3)$ spin-dependent operators (bilinear operators in the heavy quark fields and 
heavy-light operators) is included. 
Renormalization group improved expressions for the Wilson coefficients associated to these operators are obtained with LL accuracy. The computation is 
done in Coulomb gauge.

The paper is organized as follows. In Sec. \ref{Sec:Lag} we introduce the HQET Lagrangian up to $\mathcal{O}(1/m^3)$ including heavy-light 
operators (at $\mathcal{O}(1/m^3)$ only spin-dependent heavy-light operators are included. See Ref. \cite{Balzereit:1998am} for a complete basis and 
Ref. \cite{chromopolarizabilities} for relevant spin-independent operators that get LL running). The Sec. \ref{Sec:RGE} is dedicated to the 
computation of the anomalous dimensions and it is divided in three subsections: In Sec. \ref{Sec:LLlight} and Sec. \ref{sec:RGeq} the 
renormalization group equations (RGEs) for the Wilson 
coefficients associated to $1/m^3$ heavy-light operators and to $1/m^3$ bilinear operators are presented, respectively. In Sec. \ref{Sec:RGEPH} 
physical combinations are sought, and their associated RGE equations are presented. The solution of the RGEs for the physical quantities, as well as its 
numerical analysis, is displayed in Sec. \ref{Sec:numerics}. We conclude in Sec. \ref{sec:conclusions}. Finally, some necessary Feynman rules are summarized in Sec. \ref{sec:fr}.

\section{HQET Lagrangian}
\label{Sec:Lag}

The starting point is the HQET Lagrangian up to $\mathcal{O}(1/m^3)$ for a quark of mass $ m \gg \lQ$ in the rest frame, $v^\mu=(1,{\bf 0})$. It 
is given in Refs. \cite{Manohar:1997qy,Balzereit:1998am}, and reads:
\bea
&& 
{\cal L}_{\rm HQET}={\cal L}_g+{\cal L}_{Q} + {\cal L}_l
\,,
\label{LagHQET}
\\
\nn
\\
&&
{\cal L}_g=-\frac{1}{4}G^{\mu\nu \, a}G_{\mu \nu}^a +
c_1^{g}\frac{g}{4m^2} 
 f^{abc} G_{\mu\nu}^a G^{\mu \, b}{}_\alpha G^{\nu\alpha\, c}+
{\cal O}\left(\frac{1}{m^4}\right),
\label{Lg}
\\
\nn
\\
&&
\nn
{\cal L}_{Q}=
Q^{\dagger} \Biggl\{ i D_0 + \frac{c_k}{ 2 m} {\bf D}^2 
+ \frac{c_F }{ 2 m} {\bfsigma \cdot g{\bf B}} 
\\
&& \nn
+\frac { c_D}{ 8 m^2} \left({\bf D} \cdot g{\bf E} - g{\bf E} \cdot {\bf D} \right) + i \, \frac{ c_S}{ 8 m^2} 
{\bfsigma \cdot \left({\bf D} \times g{\bf E} -g{\bf E} \times {\bf D}\right) }
\\
&& \nn
+\frac {c_4 }{ 8 m^3} {\bf D}^4 + i c_M\, g { {\bf D\cdot \left[D \times B
\right] + \left[D \times B \right]\cdot D} \over 8 m^3}
+ c_{A_1}\, {g^2}\, {{\bf B}^2-{\bf E}^2 \over 8 m^3}- 
c_{A_2} { {g^2}{\bf E}^2 \over 16 m^3} 
\\ 
&& + c_{W_1}\, g { \left\{ {\bf D^2,\bfsigma
\cdot B }\right\}\over 8 m^3} - c_{W_2}\, g { {\bf D}^i\, {\bf \bfsigma
\cdot B} \, {\bf D}^i  \over 4 m^3}+ c_{p'p}\, g { {\bf \bfsigma \cdot D\, B \cdot D + D
\cdot B\, \bfsigma \cdot D}\over 8 m^3} 
\nn
\\
&&
+ c_{A_3}\, {g^2}\, \frac{1}{N_c}{\rm Tr}\left({{\bf B}^2-{\bf E}^2 \over 8 m^3}\right)- 
c_{A_4}\, {g^2}\, \frac{1}{N_c}{\rm Tr}\left({{\bf E}^2 \over 16 m^3}\right) 
\nn
\\
&&
+ i c_{B_1}\, g^2\, { {\bf \bfsigma \cdot \left(B
\times B - E \times E \right)} \over 8 m^3}
- i c_{B_2}\, g^2\, { {\bf \bfsigma \cdot \left( E \times E \right)} \over 8 m^3}
\Biggr\} Q+
{\cal O}\left(\frac{1}{m^4}\right)
\,.
\label{Lhh}
\eea
Where $Q$ is the nonrelativistic fermion field represented by a Pauli spinor. We define
$i D^0=i\partial^0 -gA^{0\,a}T^a$ and $i{\bf D}=i\bfnabla+g{\bf A}^a T^a$, where $A^{0\,a}$ and ${\bf A}^a$ represent the longitudinal and tranverse gluon fields, respectively. 
The chromoelectric field is defined as ${\bf E}^i = G^{i0}$, whereas the chromomagnetic field as ${\bf B}^i = -\eps^{ijk}G^{jk}/2$, where $\eps^{ijk}$ is
the three-dimensional totally antisymmetric tensor\footnote{
In dimensional regularization several prescriptions are possible for the $\eps^{ijk}$ tensors and $\bfsigma$, and the same prescription 
as for the calculation of the Wilson coefficients must be used.},
with $\eps^{123}=1$. The components of the vector $\bfsigma$ are the Pauli matrices. Note also the rescaling by a factor $1/N_c$ of the 
coefficients $c_{A_{3,4}}$ following Ref. \cite{chromopolarizabilities}, as compared to the definitions given in Ref. \cite{Manohar:1997qy}.

The inclusion of $n_f$ massless fermions adds an extra contribution to the HQET Lagrangian with the following structure:
\bea
&&
{\cal L}_l = \sum_{i=1}^{n_f} \bar q_i i \dsl q_i 
+\frac{\delta {\cal L}^{(2)}_q}{m^2}
+\frac{\delta {\cal L}^{(2)}_{Q q}}{m^2}
+\frac{\delta {\cal L}^{(3)}_{ q}}{m^3}
+\frac{\delta {\cal L}^{(3)}_{Q q}}{m^3} +
{\cal O}\left(\frac{1}{m^4}\right)
.
\eea
The complete set of operators at ${\cal O}(1/m^2)$ can be found in \cite{Bauer:1997gs}. They read
\bea
&& 
\delta {\cal L}^{(2)}_{Qq}
=\frac{c_1^{hl} }{8}\, g^2 \,\sum_{i=1}^{n_f}Q^{\dagger} T^a Q   \bar{q}_i\gamma^0 T^a q_i 
-\frac{c_2^{hl}}{8}\, g^2 \,\sum_{i=1}^{n_f}Q^{\dagger}\bfsigma^j T^a Q  \bar{q}_i\gamma^j\gamma_5 T^a q_i 
\nn
\\
&& \qquad\qquad
+\frac{c_3^{hl}}{ 8}\, g^2 \,\sum_{i=1}^{n_f}Q^{\dagger} Q  \bar{q}_i\gamma^0 q_i
-\frac{c_4^{hl}}{ 8}\, g^2 \,\sum_{i=1}^{n_f}Q^{\dagger}\bfsigma^j Q  \bar{q}_i\gamma^j\gamma_5 q_i,
\label{Lhl}
\\
&&
\delta {\cal L}^{(2)}_q=\frac{c_D^{l}}{4} \bar q_i \gamma_{\nu} D_{\mu} G^{\mu \nu}
 q_i 
\nn
\\
&& \qquad\qquad
+\frac {c_1^{ll}}{ 8} \, g^2 \,
\sum_{i,j =1}^{n_f} \bar{q_i} T^a \gamma^\mu q_i \ \bar{q}_j T^a \gamma_\mu q_j  
+\frac{c_2^{ll}}{ 8} \, g^2 \,
\sum_{i,j=1}^{n_f}\bar{q_i} T^a \gamma^\mu \gamma_5 q_i \ \bar{q}_j T^a \gamma_\mu \gamma_5 q_j 
\nn
\\ 
&& \qquad\qquad
+ \frac{c_3^{ll} }{ 8}\, g^2 \,
\sum_{i,j=1}^{n_f} \bar{q_i}  \gamma^\mu q_i \ \bar{q}_j \gamma_\mu q_j 
+ \frac{c_4^{ll}}{ 8}  \, g^2 \,
\sum_{i,j=1}^{n_f}\bar{q_i} \gamma^\mu \gamma_5  q_i \ \bar{q}_j \gamma_\mu \gamma_5 q_j.
\label{Ll}
\eea

However, the light-light operators $\delta\mathcal{L}_q^{(2)}$ and $\delta\mathcal{L}_q^{(3)}$, as well as the $1/m^2$ gluonic operator with associated
Wilson coefficient $c_1^g$, contribute at NLL or beyond, so we will not consider them any further.

The $\mathcal{O}(1/m^3)$ (dimension 7) heavy-light operators were considered in detail in Ref. \cite{Balzereit:1998am} and they can be 
found in Eq. (10) of that reference. However, we will not consider all of them, but only those that get LL running and that could affect the 
LL running of $c_{p'p}$, $c_{W_1}$, $c_{W_2}$, $c_{B_1}$ and $c_{B_2}$. Initially, we can disregard some of them because of its spin independence 
or just by using the heavy quark equation of motion. After that, we face the following operators:

\begin{equation}
 \mathcal{M}_{3\pm}^{(3h)s/o} = \pm g_s^2[\bar q_l \gamma_\mu \mathcal{C}_{s/o}^a q_l][\bar h_v i\sigma^{\mu\nu}\mathcal{C}_{s/o}^a i D_\nu^{\pm}h_v]\,,
\end{equation}

\begin{equation}
 \mathcal{M}_{5\pm}^{(3h)s/o} = \pm g_s^2[\bar q_l i\sigma_{\mu\lambda}v^\lambda \mathcal{C}_{s/o}^a q_l]
 [\bar h_v i\sigma^{\mu\nu}\mathcal{C}_{s/o}^a i D_\nu^{\pm}h_v]\,,
\end{equation}

\begin{equation}
 \mathcal{M}_{7\pm}^{(3h)s/o} = \pm g_s^2[\bar q_l \gamma_5 \slashed v \mathcal{C}_{s/o}^a q_l][\bar h_v \gamma_5 \mathcal{C}_{s/o}^a i \slashed D^{\pm}h_v]\,,
\end{equation}

\begin{equation}
 \mathcal{M}_{9\pm}^{(3h)s/o} = \pm g_s^2[\bar q_l \gamma_5 \mathcal{C}_{s/o}^a q_l][\bar h_v \gamma_5 \mathcal{C}_{s/o}^a i \slashed D^{\pm}h_v]\,,
\end{equation}

\begin{equation}
\mathcal{M}_{5\pm}^{(3l)s/o} = \pm g_s^2[\bar q_l i \sigma^{\mu\nu}\mathcal{C}_{s/o}^a (ivD^{\pm})q_l][\bar h_v i\sigma_{\mu\nu}\mathcal{C}_{s/o}^a h_v]\,,
\end{equation}

\begin{equation}
\mathcal{M}_{6\pm}^{(3l)s/o} = \pm g_s^2[\bar q_l \gamma_5 \slashed v \mathcal{C}_{s/o}^a iD_\mu^{\pm}q_l]
[\bar h_v \gamma_5 \gamma^\mu \mathcal{C}_{s/o}^a h_v]\,,
\end{equation}

\begin{equation}
\mathcal{M}_{7\pm}^{(3l)s/o} = \pm g_s^2[\bar q_l \gamma_5 \gamma^\mu \mathcal{C}_{s/o}^a (ivD^{\pm})q_l]
[\bar h_v \gamma_5 \gamma_\mu \mathcal{C}_{s/o}^a h_v]\,,
\end{equation}

\begin{equation}
\mathcal{M}_{8\pm}^{(3l)s/o} = \pm g_s^2[\bar q_l \gamma_5 \mathcal{C}_{s/o}^a iD_\mu^{\pm}q_l][\bar h_v \gamma_5 \gamma^\mu \mathcal{C}_{s/o}^a h_v]\,,
\end{equation}

\begin{equation}
\mathcal{M}_{10\pm}^{(3l)s/o} = \pm g_s^2[\bar q_l \gamma_\nu \mathcal{C}_{s/o}^a iD_\mu^{\pm}q_l][\bar h_v i\sigma^{\mu\nu} \mathcal{C}_{s/o}^a h_v]\,.
\end{equation}
Where $iD_\mu^+=i \stackrel{\rightarrow}{\partial}_\mu - g A_\mu^a T^a$ and $iD_\mu^-=i\stackrel{\leftarrow}{\partial}_\mu +gA_\mu^a T^a$, meaning the 
arrows over the derivatives that they act over fields in the left/right hand depending on the direction of the arrow (they only act over heavy quark fields 
or over light quark fields), $\mathcal{C}_s^a=1$ and $\mathcal{C}_o^a=T^a$ and $\sigma^{\mu\nu}=\frac{i}{2}[\gamma^\mu,\gamma^\nu]$. In our case, we work 
in the rest frame, so that $v^\mu=(1,\bf 0)$ and $h_v \equiv Q$. It is also understood that in the octet case the covariant derivative 
stands left/right of the color matrix when acting to the left/right. Moreover, we are in the heavy-quark sector, and not in the antiquark one, so we 
can project to this sector. Note that we have not displayed the operator $\mathcal{M}_{9\pm}^{(3l)s/o}$ because it is 
wrong (there are typographic mistakes and even free indices) and should be corrected. Fortunately, as 
we will see later on, this operator is not relevant for the computation of the LL running of the Wilson coefficients, since the operators that are left 
are enough to absorve all divergences coming from one-loop diagrams. After all these simplifications, the previous operators can be written as:

\begin{equation}
\label{lfo1}
 \mathcal{M}_{3\pm}^{(3h)s/o} = \mp g_s^2[\bar q_l \gamma^i \mathcal{C}_{s/o}^a q_l]
 [Q^\dagger i\epsilon^{ijk}\bfsigma^k \mathcal{C}_{s/o}^a i {\bf D}^{j\,\pm} Q]\,,
\end{equation}

\begin{equation}
\label{lfo2}
 \mathcal{M}_{5\pm}^{(3h)s/o} = \pm g_s^2[\bar q_l \gamma^i \gamma^0 \mathcal{C}_{s/o}^a q_l]
 [Q^\dagger i\epsilon^{ijk}\bfsigma^k \mathcal{C}_{s/o}^a i {\bf D}^{j\,\pm} Q]\,,
\end{equation}

\begin{equation}
\label{lfo3}
 \mathcal{M}_{7\pm}^{(3h)s/o} = \mp g_s^2[\bar q_l \gamma_5 \gamma^0 \mathcal{C}_{s/o}^a q_l]
 [ Q^\dagger \bfsigma^i \mathcal{C}_{s/o}^a i  {\bf D}^{i\,\pm} Q]\,,
\end{equation}

\begin{equation}
\label{lfo4}
 \mathcal{M}_{9\pm}^{(3h)s/o} = \mp g_s^2[\bar q_l \gamma_5 \mathcal{C}_{s/o}^a q_l]
 [Q^\dagger \bfsigma^i  \mathcal{C}_{s/o}^a i {\bf D}^{i\,\pm} Q]\,,
\end{equation}

\begin{equation}
\label{lfo5}
\mathcal{M}_{5\pm}^{(3l)s/o} = \mp g_s^2[\bar q_l \gamma^i \gamma^j\mathcal{C}_{s/o}^a i D_0^{\pm} q_l]
[Q^\dagger i \epsilon^{ijk}\bfsigma^k \mathcal{C}_{s/o}^a Q]\,,
\end{equation}

\begin{equation}
\label{lfo7}
\mathcal{M}_{6\pm}^{(3l)s/o} = \mp g_s^2[\bar q_l \gamma_5 \gamma^0 \mathcal{C}_{s/o}^a i{\bf D}^{i\,\pm}q_l]
[Q^\dagger \bfsigma^i  \mathcal{C}_{s/o}^a Q]\,,
\end{equation}

\begin{equation}
\label{lfo8}
\mathcal{M}_{7\pm}^{(3l)s/o} = \pm g_s^2[\bar q_l \gamma_5 \gamma^i \mathcal{C}_{s/o}^a iD_0^{\pm} q_l]
[Q^\dagger \bfsigma^i  \mathcal{C}_{s/o}^a Q]\,,
\end{equation}

\begin{equation}
\label{lfo9}
\mathcal{M}_{8\pm}^{(3l)s/o} = \mp g_s^2[\bar q_l \gamma_5 \mathcal{C}_{s/o}^a i{\bf D}^{i\,\pm}q_l]
[Q^\dagger \bfsigma^i \mathcal{C}_{s/o}^a Q]\,,
\end{equation}

\begin{equation}
\label{lfo10}
\mathcal{M}_{10\pm}^{(3l)s/o} = \mp g_s^2[\bar q_l \gamma^j \mathcal{C}_{s/o}^a i{\bf D}^{i\,\pm}q_l]
[Q^\dagger i \epsilon^{ijk}\bfsigma^k \mathcal{C}_{s/o}^a Q]\,.
\end{equation}
We then have
\be
\delta {\cal L}^{(3)}_{Qq}= \sum_{l=1}^{n_f}\sum_m d_m^{hl}\mathcal{O}_m
\,,\ee
where the $\mathcal{O}_m$ operators are all the possible linear independent combinations of the $\mathcal{M}$ operators. In the present article, only those 
linear combinations whose associated Wilson coefficients get LL running will be defined. The discussion is reserved to Sec.~\ref{Sec:LLlight}.

\section{Anomalous dimensions for $1/m^3$ spin-dependent operators}
\label{Sec:RGE}

In this section, the anomalous dimensions of the Wilson coefficients associated to the $1/m^3$ spin-dependent operators is computed 
at $\mathcal{O}(\alpha)$. On the one hand, for the operators bilinear in the heavy quark fields, the anomalous dimensions 
in the case of $n_f=0$ were already computed in Ref. \cite{lmp}, so only the contribution  from heavy-light operators remains to be computed. On 
the other hand, for the heavy-light operators, all the contributions to their anomalous dimensions must be computed, the one coming from the bilinear sector and the one coming from 
the heavy-light sector. In the former, the anomalous dimensions are determined through the scattering, at one loop order, of a heavy quark with a 
transverse gluon, whereas in the later, the anomalous dimensions are determined through the scattering, at one loop order, of a heavy quark with 
a light quark. We follow the procedure used in Refs. \cite{chromopolarizabilities,lmp}, in which a minimal basis of operators is considered, so the computation resembles 
the one of an S-matrix element, and both reducible and irreducible diagrams must be considered. Since we are only interested in the anomalous dimensions, 
it is enough to determine the UV pole of the integrals. The computation is organized in powers of $1/m$, up to ${\cal O}(1/m^3)$, by 
considering all possible insertions of the HQET Lagrangian operators. In general, external particles will be considered to be on-shell i.e. free 
asymptotic states, so the free equations of motion (EOM) will be used throughout. The computation is done in the Coulomb gauge and in dimensional 
regularization.

It is important to recall that the Compton scattering analysis at $\mathcal{O}(1/m^3)$ in Ref. \cite{lmp} showed that 
$\bar c_{W} = c_{W_1}-c_{W_2}$, $\bar c_{B_1} = c_{B_1}-2c_{W_1}$, $c_{B_2}$  and $c_{p'p}$ are physical combinations i.e. they are gauge independent. This observation will 
be crucial in order to determine what combinations of Wilson coefficients associated to heavy-light operators will be gauge independent.

The Wilson coefficients of the kinetic terms will be kept explicit for tracking purposes even though they are 
protected by reparametrization invariance i.e. $c_k=c_4=1$ to any order in perturbation theory \cite{Luke:1992cs}. The Wilson coefficient 
$c_{p'p}=c_F-1$ and the physical combination $\bar c_W=1$ are fixed by reparametrization 
invariance, as well \cite{Manohar:1997qy}. We will check by explicit calculation that these relations are satisfied at LL even adding massless quarks.

For the aimed calculation only the renormalization of the heavy quark field, massless quark field and the strong coupling $g$, in the 
Coulomb gauge, are needed. They read (We define $D=4+2\epsilon$ as the number of space-time dimensions. The number of spatial dimensions 
is $d=3+2\epsilon$, whereas there is only one temporal dimension):
\bea
\nn
Z_g&=&1+\frac{11}{6}C_A\frac{\al}{4\pi}\frac{1}{\epsilon} -\frac{2}{3}T_Fn_f\frac{\al}{4\pi}\frac{1}{\epsilon}
\,,
\quad 
Z_l=1+C_F\frac{\al}{4\pi}\frac{1}{\epsilon}
\,, 
\quad
Z_h=1+\frac{4}{3}C_F\frac{{\bf p}^2}{m^2}\frac{\al}{4\pi}\frac{1}{\epsilon}
\,,
\eea
where
\begin{eqnarray}
  \cf=\frac{N_c^2-1}{2N_c}=\frac{4}{3}\,, \qquad C_A = N_c=3\,, \qquad T_F = \frac{1}{2}\,.
\end{eqnarray}
 
\subsection{$1/m^3$ heavy-light operators: LL running of $d_i^{hl}$}
\label{Sec:LLlight}

The Wilson coefficients associated to the heavy-light operators $c_i^{hl}$ and $d_i^{hl}$ evaluated at the hard scale are of $\mathcal{O}(\alpha)$ 
(so at the order of interest, i.e. at tree level, the maching condition is zero). This is so because, such operators, can not be generated at tree level in 
the underlying theory, QCD. Given this condition, the only way they can get LL running is through mixing with other Wilson coefficients that 
get LL running.

\begin{figure}[!htb]
	\includegraphics[width=1.00\textwidth]{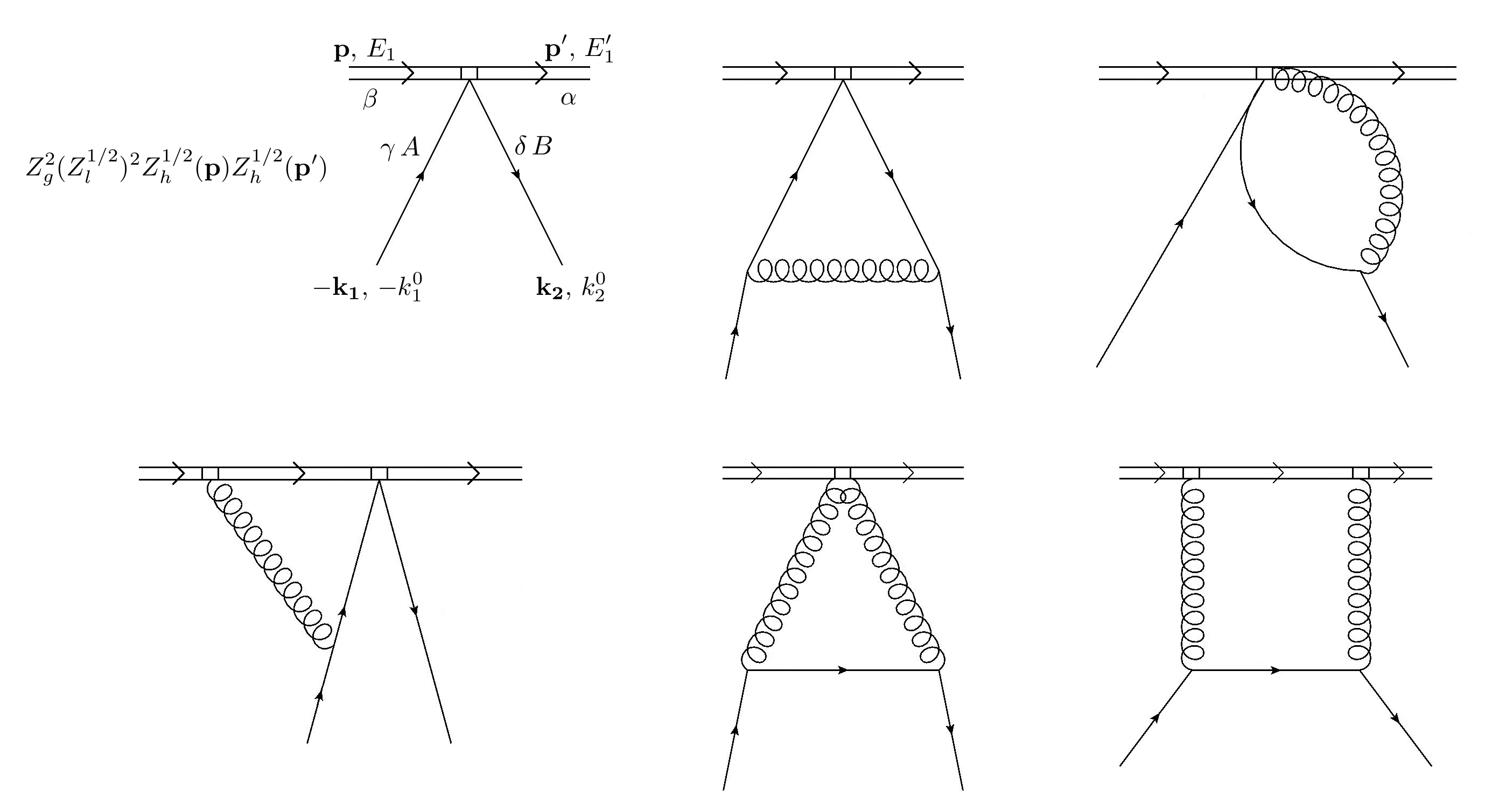}
\caption{Topologies contributing to the LL running of Wilson coefficients associated to $1/m^3$ spin-dependent heavy-light operators. The first diagram is 
the tree level diagram multiplied by the renormalization of the external fields and coupling. The other diagrams are the one-loop topologies that also 
contribute. In general the depicted gluon can be either longitudinal or transverse. All possible vertices and 
insertions with the right counting in $1/m$ should be considered to generate the diagrams.}
\label{PlotsHeavyLight}   
\end{figure}

In order to determine which operators of Eqs. (\ref{lfo1})-(\ref{lfo10}) are relevant i.e. which operators get LL running, we compute the 
scattering of a heavy quark 
with a light quark. The Wilson coefficients associated to these operators will get LL running if there is mixing with the Wilson coefficients of the bilinear sector 
up to $\mathcal{O}(1/m^3)$ or with the Wilson coefficients associated to heavy-light operators up to $\mathcal{O}(1/m^2)$. Finally, one also has 
to compute the self-running with the Wilson coefficients associated to the $\mathcal{O}(1/m^3)$ heavy-light operators. However, the later are not relevant to determine if the 
Wilson coefficients get LL running or not. To the order of interest, the scattering must be computed at one loop. Divergences coming from Feynman diagrams 
will be absorbed in the Wilson coefficients $d_i^{hl}$ determining its running. What we find is what we already advanced in previous sections, not all the 
operators in Eqs. (\ref{lfo1})-(\ref{lfo10}) get LL running, but only a combination of some of them. In particular, there are eight different operators 
relevant for our discussion, which read

\begin{equation}
 \label{O4}
 \mathcal{O}_4 = \mathcal{M}_{7+}^{(3h)o} + \mathcal{M}_{7-}^{(3h)o}\,,
\end{equation}

\begin{equation}
\label{O5}
 \mathcal{O}_5 = \mathcal{M}_{7+}^{(3h)s} + \mathcal{M}_{7-}^{(3h)s}\,,
\end{equation}

\begin{equation}
\label{O6}
 \mathcal{O}_6 = \mathcal{M}_{6+}^{(3l)o} + \mathcal{M}_{6-}^{(3l)o}\,,
\end{equation}

\begin{equation}
\label{O7}
 \mathcal{O}_{7} = \mathcal{M}_{6+}^{(3l)s} + \mathcal{M}_{6-}^{(3l)s}\,,
\end{equation}

\begin{equation}
\label{O8}
 \mathcal{O}_{8} = \mathcal{M}_{7+}^{(3l)o} + \mathcal{M}_{7-}^{(3l)o}\,,
\end{equation}

\begin{equation}
\label{O9}
 \mathcal{O}_{9} = \mathcal{M}_{7+}^{(3l)s} + \mathcal{M}_{7-}^{(3l)s}\,,
\end{equation}

\begin{equation}
\label{O10}
 \mathcal{O}_{10} = \mathcal{M}_{10+}^{(3l)o} - \mathcal{M}_{10-}^{(3l)o}\,,
\end{equation}

\begin{equation}
\label{O11}
 \mathcal{O}_{11} = \mathcal{M}_{10+}^{(3l)s} - \mathcal{M}_{10-}^{(3l)s}\,.
\end{equation}
The Feynman rules associated to these operators are displayed in App. \ref{sec:fr}. The running of these operators is obtained from the 
diagrams (topologies) drawn in Fig.~\ref{PlotsHeavyLight}. They produce around 67 diagrams to be computed without 
counting crossed and inverted ones.

The RGE we obtain are 

$$\nu \frac{d}{d\nu} d_{4}^{hl} = \frac{\alpha}{\pi}\bigg[
(8C_F-3C_A)\bigg(\frac{1}{32} c_{W_1} 
- \frac{1}{32} c_{W_2}
+ \frac{1}{16} c_{p'p}
+ \frac{1}{64} c_S c_k$$
\begin{equation}
 - \frac{1}{32} c_S c_F 
- \frac{5}{32}c_F c_k^2
+ \frac{5}{64} c_F^2 c_k \bigg)
- \frac{1}{4}d_4^{hl}(3C_A - 2\beta_0)\bigg]\,,
\end{equation}\\

$$\nu \frac{d}{d\nu} d_{5}^{hl} = \frac{\alpha}{\pi}\bigg[
C_F(2C_F-C_A)\bigg( - \frac{1}{16}c_{W_1}
+ \frac{1}{16} c_{W_2}
- \frac{1}{8}c_{p'p}
- \frac{1}{32} c_S c_k$$
\begin{equation}
+ \frac{1}{16} c_S c_F 
+ \frac{5}{16} c_F c_k^2 
- \frac{5}{32} c_F^2 c_k  \bigg) 
+ \frac{1}{2} d_5^{hl} \beta_0 \bigg]\,,
\end{equation}\\

$$\nu \frac{d}{d\nu} d_{6}^{hl} = \frac{\alpha}{\pi}\bigg[
  \frac{1}{192} c_{B_1}C_A 
  + \frac{1}{192} c_S c_F C_A  
 - \frac{5}{96} c_F^2 c_k C_A  
 + \frac{1}{64} c_2^{hl} c_F (8C_F-3C_A)
 + \frac{1}{16} c_4^{hl} c_F$$
\begin{equation}
 + \frac{1}{3} d_4^{hl}(8C_F-3C_A) 
+ \frac{4}{3} d_5^{hl} 
+ \frac{1}{6} d_6^{hl}(5C_F-5C_A+3\beta_0)
 + \frac{5}{12} d_8^{hl}(2C_F-C_A)
 + \frac{1}{12} d_{10}^{hl}C_A \bigg]\,,
\end{equation}\\

\begin{equation}
\nu \frac{d}{d\nu} d_{7}^{hl} =  
\frac{\alpha}{\pi}\bigg[ 
-C_F(2C_F - C_A)\bigg(  \frac{1}{32} c_2^{hl} c_F 
+ \frac{2}{3} d_4^{hl} \bigg)
+ \frac{1}{6} d_7^{hl}(5C_F+3\beta_0)
+ \frac{5}{6} d_9^{hl}C_F \bigg]\,,
\end{equation}\\

$$\nu \frac{d}{d\nu} d_{8}^{hl} = 
\frac{\alpha}{\pi} \bigg[ -\frac{1}{32} c_{W_1}C_A 
- \frac{1}{192} c_{B_1}C_A   
- \frac{1}{96} c_{B_2}C_A $$
$$- \frac{1}{64} c_D c_F C_A 
-\frac{1}{64} c_S c_k C_A
+ \frac{1}{192} c_S c_F C_A 
+ \frac{5}{32} c_F c_k^2 C_A 
- \frac{5}{96} c_F^2 c_k C_A $$
$$- \frac{1}{64} c_1^{hl} c_F C_A 
- \frac{1}{16} c_2^{hl} c_k (8C_F-3C_A)
+ \frac{1}{32}c_2^{hl} c_F (8C_F-3C_A)
- \frac{1}{4} c_4^{hl} c_k
+ \frac{1}{8} c_4^{hl} c_F $$
\begin{equation}
- \frac{1}{12} d_4^{hl}(8C_F-3C_A) 
- \frac{1}{3} d_5^{hl} 
+ \frac{1}{6} d_6^{hl}(3C_F- 2C_A)
+ \frac{1}{2} d_8^{hl}(C_F-2C_A+\beta_0)
+ \frac{1}{6} d_{10}^{hl}C_A\bigg],\,
\end{equation}\\

\begin{equation}
\nu \frac{d}{d\nu} d_{9}^{hl} = 
 \frac{\alpha}{\pi} \bigg[ 
 C_F(2C_F-C_A)\bigg(\frac{1}{8} c_2^{hl} c_k 
 - \frac{1}{16} c_2^{hl} c_F
+ \frac{1}{6} d_4^{hl} \bigg)
+ \frac{1}{2} d_7^{hl}C_F
+ \frac{1}{2} d_9^{hl}(C_F+\beta_0) \bigg],\,
\end{equation}\\

$$\nu \frac{d}{d\nu} d_{10}^{hl} = \frac{\alpha}{\pi}\bigg[
\frac{1}{48} c_{W_1}C_A  
+ \frac{1}{192} c_{W_2}C_A
- \frac{7}{192} c_{B_1} C_A
- \frac{1}{48} c_{B_2}C_A 
+ \frac{1}{384} c_{p'p} C_A $$
$$+ \frac{1}{128} c_D c_F C_A 
+ \frac{1}{96} c_S c_k C_A  
- \frac{7}{384} c_S c_F C_A  
 + \frac{5}{64} c_F^2 c_k C_A  $$
$$+ \frac{1}{128} c_1^{hl} c_F C_A 
- \frac{1}{128} c_2^{hl} c_F(8C_F-3C_A)
 - \frac{1}{32} c_4^{hl} c_F $$
$$+ \frac{1}{24} d_4^{hl}(8C_F-3C_A)
+ \frac{1}{6} d_5^{hl}
- \frac{1}{24} d_6^{hl}(4C_F-3C_A)$$
\begin{equation}
 - \frac{1}{12} d_8^{hl}(2C_F-C_A)
 - \frac{1}{24} d_{10}^{hl}(11C_A-12\beta_0) \bigg]\,,
\end{equation}\\

\begin{equation}
\nu \frac{d}{d\nu} d_{11}^{hl} = \frac{\alpha}{\pi} \bigg[
C_F(2C_F - C_A) \bigg( \frac{1}{64} c_2^{hl} c_F 
- \frac{1}{12} d_4^{hl}\bigg)
- \frac{1}{6} d_7^{hl}C_F
-\frac{1}{6} d_9^{hl}C_F 
+ \frac{1}{2} d_{11}^{hl}\beta_0  \bigg]\,.
\end{equation}
The RGE of the remaining Wilson coefficients $d_i^{hl}$ have the structure ($i,j > 11$)
\be
\nu\frac{d}{d\nu}d_i^{hl}=\frac{\al}{\pi}A_{ij}d_j^{hl}
\,.
\ee
And for this reason, they are NLL.

\subsection{$1/m^3$ heavy quark bilinear operators: LL running of $c_{p'p}$, $c_{Wi}$ and $c_{Bi}$}
\label{sec:RGeq}

\begin{figure}[!htb]
	\includegraphics[width=0.95\textwidth]{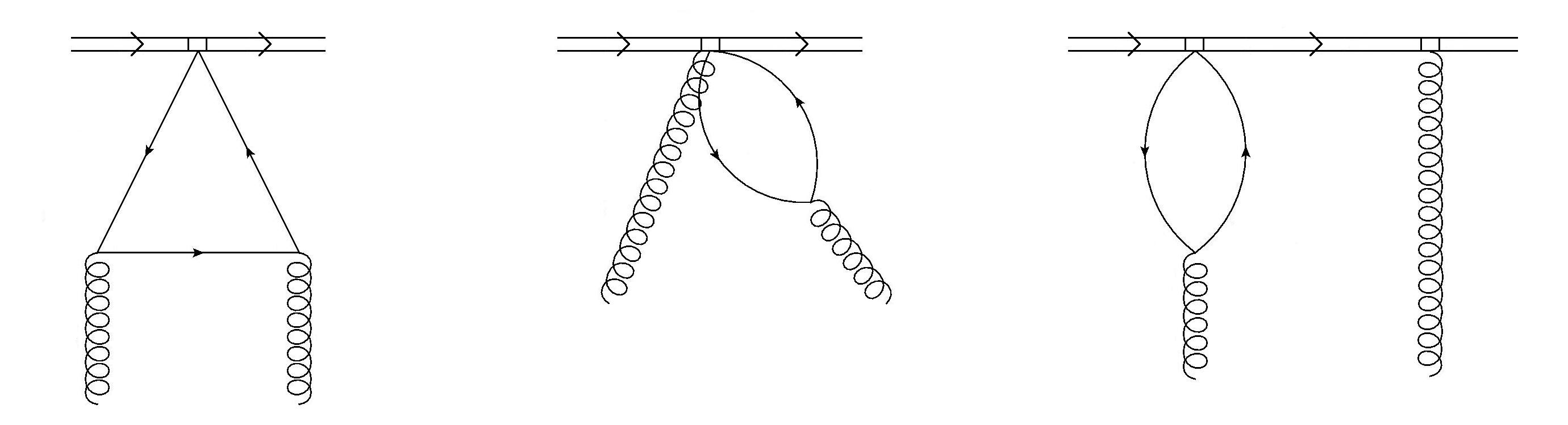}
\caption{One loop topologies contributing to the anomalous dimensions of the Wilson 
coefficients of the $1/m^3$ operators bilinear in the heavy quark field. All diagrams are generated from these topologies by 
considering all possible vertices and kinetic insertions up to ${\cal O}(1/m^3)$.
\label{PlotsHeavygluon1}}   
\end{figure}

Let's consider the $1/m^3$ spin-dependent operators bilinear in the heavy quark field of the HQET Lagrangian, namely, the running of the unphysical 
set: $\{c_{W_1},c_{W_2},c_{B_1},c_{B_2}, c_{p'p}\}$. The most difficult part of the work was already done in Ref. \cite{lmp}. 
The only part which is left is the contribution due to heavy-light operators i.e. the running of these Wilson coefficients with $c_i^{hl},\,i=1,\ldots,4$ 
and $d_i^{hl},\, i=4,\ldots,11$. The procedure we use is the same that in Refs. \cite{chromopolarizabilities,lmp}. We compute the elastic 
scattering of a heavy quark with a transverse gluon only considering diagrams involving the vertices coming from $1/m^2$ and $1/m^3$ spin-dependent 
heavy-light operators. Diagrams are constructed from the topologies shown in Fig.~\ref{PlotsHeavygluon1} by considering all possible 
vertices and kinetic insertions to the appropriate order in $1/m$. Note that diagrams of lower order than $1/m^3$ also must 
be considered, as the use of the heavy quark EOM, $E= c_k \frac{{\bf p}^2}{2m}$, adds extra powers of $1/m$ 
in those terms which are proportional to the energy. The 
topologies drawn in Fig.~\ref{PlotsHeavygluon1} generate around 21 diagrams without taking into account permutations and crossing.
The RGEs for the unphysical set $\{c_{W_1},c_{W_2},c_{B_1},c_{B_2}, c_{p'p}\}$, in Coulomb gauge, read

\begin{equation}
\label{rgecpp}
 \nu \frac{d}{d\nu}c_{p'p} = \gamma_{c_{p'p},\, Q^\dagger Q}\,,
\end{equation}

\begin{equation}
\label{rgecW1}
 \nu \frac{d}{d\nu}c_{W_1} = \gamma_{c_{W_1},\, Q^\dagger Q}
 - \frac{\alpha}{\pi}  T_F n_f\bigg(  \frac{8}{3}d_6^{hl}
 - \frac{8}{3} d_8^{hl} 
 + \frac{16}{3} d_{10}^{hl} \bigg) \,,
\end{equation}

\begin{equation}
\label{rgecW2}
 \nu \frac{d}{d\nu}c_{W_2} = \gamma_{c_{W_2},\, Q^\dagger Q} 
- \frac{\alpha}{\pi} T_F n_f  \bigg(  \frac{8}{3}d_6^{hl}
- \frac{8}{3} d_8^{hl}
+ \frac{16}{3} d_{10}^{hl}\bigg)\,,
\end{equation}

\begin{equation}
\label{rgecB1}
 \nu \frac{d}{d\nu} c_{B_1} = \gamma_{c_{B_1},\, Q^\dagger Q}  
 - \frac{\alpha}{\pi}T_F n_f  \bigg(  \frac{8}{3}d_6^{hl}
 - 8 d_8^{hl}  
 + \frac{32}{3} d_{10}^{hl}\bigg)\,,
\end{equation}

\begin{equation}
 \nu \frac{d}{d\nu} c_{B_2} = \gamma_{c_{B_2},\, Q^\dagger Q}  
-  \frac{\alpha}{\pi} T_F n_f \bigg( \frac{16}{3}d_6^{hl} 
+ \frac{16}{3} d_8^{hl}  \bigg) \,.
\end{equation}
Where $\gamma_{c_i,\, Q^\dagger Q}$ is the anomalous dimension of the Wilson coefficient $c_i$ found in Ref. \cite{lmp}, that comes only from 
the terms of the HQET Lagrangian bilinear in the heavy quark field or, what is the same, it is the anomalous dimension for $n_f=0$. 

\subsection{LL running of physical quantities}
\label{Sec:RGEPH}

In the previous sections, Sec. \ref{Sec:LLlight} and Sec. \ref{sec:RGeq}, we found the running of the Wilson coefficients associated to 
the $1/m^3$ HQET Lagrangian operators including spectator 
quarks. However, it is well known from Ref. \cite{lmp} that Eqs. (\ref{rgecW1})-(\ref{rgecB1}) are not physical. So the next step, is 
to compute the RGEs for the the known physical quantities $\bar c_W=c_{W_1}-c_{W_2}$, $\bar c_{B_1}=c_{B_1}-2c_{W_1}$, 
$c_{B_2}$ and $c_{p'p}$. They read

\begin{equation}
\label{rgephcpp}
 \nu \frac{d}{d\nu}c_{p'p} = \gamma_{c_{p'p},\, Q^\dagger Q}\,,
\end{equation}

\begin{equation}
\label{rgephcW}
 \nu \frac{d}{d\nu}\bar c_{W} = \gamma_{\bar c_W,\, Q^\dagger Q}\,,
\end{equation}

\begin{equation}
\label{rgephcB1}
 \nu \frac{d}{d\nu} \bar c_{B_1} = \gamma_{\bar c_{B_1},\, Q^\dagger Q}
+ \frac{8}{3} \bar d_8^{hl} T_F n_f \frac{\alpha}{\pi}\,,
\end{equation}

\begin{equation}
\label{rgephcB2}
 \nu \frac{d}{d\nu} c_{B_2} = \gamma_{c_{B_2},\, Q^\dagger Q} 
- \frac{16}{3} \bar d_8^{hl} T_F n_f \frac{\alpha}{\pi}\,.
\end{equation}
Note that Eqs. (\ref{rgephcpp}-\ref{rgephcW}) satisfy the reparametrization invariant relations given in Ref. \cite{Manohar:1997qy}, even 
with the inclusion of spectator quarks. From these equations we learn that $\bar d_8^{hl} = d_6^{hl} + d_8^{hl}$ must be physical as it appears in the running of physical combinations. Indeed,
since the running of $d_6^{hl}$ and $d_8^{hl}$ can not be written in terms of gauge-independent quantities\footnote{If one assumes that $d_6^{hl}$ and 
$d_{8}^{hl}$ are gauge-independent, their RGEs can be written only in terms of $c_{W_1}$ and $d_{10}^{hl}$, which should combine in a gauge independent way. 
However, the combination in the RGEs of $d_6^{hl}$ and $d_8^{hl}$ is different making it impossible.}, $\bar d_8^{hl}$ must 
be a physical combination, whereas $d_6^{hl}$ and $d_8^{hl}$ alone are gauge dependent. The gauge independence of the RGE for $\bar d_8^{hl}$ also 
implies the existence of another physical combination, $\bar d_{10}^{hl} = 8d_6^{hl} + 8d_{10}^{hl} - c_{W_1}$, whose running also depends only on physical 
quantities, as expected. The running of these two physical combinations also depend on $d_4^{hl}$ and $d_5^{hl}$, 
which happen to be gauge independent, as their running only depend on physical quantities and on themselves, and they do not 
combine with any gauge dependent quantity in the running of gauge independent combinations. The Wilson coefficients $d_{7}^{hl}$, $d_{9}^{hl}$ and 
$d_{11}^{hl}$ do not mix with $c_{p'p}$, $\bar c_W$, $\bar c_{B_1}$, $c_{B_2}$, $d_{4}^{hl}$, $d_{5}^{hl}$, $\bar d_8^{hl}$ and $\bar d_{10}^{hl}$, so 
they are not necessary to determine their running. Since they do not appear in known physical quantities we do not dare to talk about 
their gauge dependence. The RGEs for the physical set of light fermion Wilson coefficients read

$$\nu \frac{d}{d\nu} d_{4}^{hl} = \frac{\alpha}{\pi} \bigg[ 
- \frac{1}{4}d_4^{hl}(3C_A - 2\beta_0)
+ (8C_F-3C_A)\bigg( \frac{1}{32} \bar c_{W}
+ \frac{1}{16} c_{p'p}$$
\begin{equation}
\label{rged4hl}
 + \frac{1}{64} c_S c_k
- \frac{1}{32} c_S c_F
- \frac{5}{32}c_F c_k^2 
+ \frac{5}{64} c_F^2 c_k \bigg) \bigg]\,,
\end{equation}\\

$$\nu \frac{d}{d\nu} d_{5}^{hl} = 
\frac{\alpha}{\pi} \bigg[ \frac{1}{2} d_5^{hl} \beta_0 
+ C_F(2C_F-C_A)\bigg( - \frac{1}{16}\bar c_{W}
- \frac{1}{8}c_{p'p} $$
\begin{equation}
 - \frac{1}{32} c_S c_k 
+ \frac{1}{16} c_S c_F 
+ \frac{5}{16} c_F c_k^2
- \frac{5}{32} c_F^2 c_k  \bigg) \bigg]\,,
\end{equation}\\

$$\nu \frac{d}{d\nu} \bar d_{8}^{hl} = 
 \frac{\alpha}{\pi} \bigg( - \frac{1}{96} c_{B_2}C_A
+ \frac{1}{4} d_4^{hl}(8C_F-3C_A)
+ d_5^{hl}
+ \frac{1}{12} \bar d_8^{hl}(16C_F -17C_A + 6\beta_0)$$
$$+ \frac{1}{32} \bar d_{10}^{hl} C_A
- \frac{1}{64} c_S c_k C_A  
+ \frac{1}{96} c_S c_F C_A 
+ \frac{5}{32} c_F c_k^2 C_A
- \frac{5}{48} c_F^2 c_k C_A  $$
\begin{equation}
 - \frac{1}{64} \bar c_1^{hl} c_F C_A 
- \frac{1}{16} c_2^{hl} c_k (8C_F-3C_A)
+ \frac{3}{64} c_2^{hl} c_F (8C_F-3C_A) 
- \frac{1}{4} c_4^{hl} c_k 
+ \frac{3}{16} c_4^{hl} c_F \bigg)\,,
\end{equation}\\

$$\nu \frac{d}{d\nu} \bar d_{10}^{hl} =
\frac{\alpha}{\pi} \bigg( - \frac{1}{24} c_{B_2}C_A 
+ 3 d_4^{hl}(8C_F-3C_A)
+ 12 d_5^{hl} 
+ \frac{2}{3} \bar d_8^{hl}(8C_F -15 C_A + 3\beta_0)$$
$$+ \frac{35}{24} \bar d_{10}^{hl} C_A
+ \frac{13}{24} \bar c_W C_A
- \frac{1}{48} c_{p'p} C_A  
- \frac{5}{24}c_S c_k C_A  
+ \frac{1}{16} c_S c_F C_A 
+ \frac{1}{12}c_F c_k^2 (16C_F + 15C_A)$$
\begin{equation}
- \frac{2}{3} c_F^2 c_k C_A 
+ \frac{1}{16} \bar c_1^{hl} c_F C_A 
+ \frac{1}{16} c_2^{hl} c_F(8C_F-3C_A) 
+ \frac{1}{4} c_4^{hl} c_F\bigg)\,,
\end{equation}\\

\begin{equation}
 \nu \frac{d}{d\nu} d_{7}^{hl} =  
\frac{\alpha}{\pi} \bigg( 
-\frac{2}{3} d_4^{hl}C_F(2C_F-C_A)
+ \frac{1}{6} d_7^{hl}(5C_F+3\beta_0)
+ \frac{5}{6} d_9^{hl}C_F 
-\frac{1}{32} c_2^{hl} c_F C_F(2C_F - C_A)\bigg)\,,
\end{equation}\\

$$\nu \frac{d}{d\nu} d_{9}^{hl} = 
\frac{\alpha}{\pi} \bigg( \frac{1}{6} d_4^{hl}C_F(2C_F - C_A) 
+ \frac{1}{2} d_7^{hl}C_F
+ \frac{1}{2} d_9^{hl}(C_F+\beta_0)$$
\begin{equation}
 + \frac{1}{8} c_2^{hl} c_k C_F(2C_F-C_A) 
- \frac{1}{16} c_2^{hl} c_F C_F(2C_F - C_A)\bigg)\,,
\end{equation}\\

\begin{equation}
\label{rged11hl}
 \nu \frac{d}{d\nu} d_{11}^{hl} = \frac{\alpha}{\pi}  \bigg(
- \frac{1}{12} d_4^{hl}C_F(2C_F-C_A) 
- \frac{1}{6} d_7^{hl}C_F 
-\frac{1}{6} d_9^{hl}C_F
+ \frac{1}{2} d_{11}^{hl}\beta_0
+ \frac{1}{64} c_2^{hl} c_F C_F(2C_F - C_A)\bigg)\,.
\end{equation}
Note that we include the Wilson coefficients $d_7^{hl}$, $d_9^{hl}$ and $d_{11}^{hl}$ despite of we do not know if they are physical or not. We do 
so because, we will solve also these RGEs in the next section, as it can be useful in the future. It it quite remarkable that the RGEs depend only 
on gauge-independent combinations of Wilson coefficients: $\bar c_{W}$, $\bar c_{B_1}$, 
$\bar d_{8}^{hl}$, $\bar d_{10}^{hl}$ and $\bar c_1^{hl}=c_D + c_1^{hl}$ (see Refs. \cite{Pineda:2001ra,chromopolarizabilities} for discussions 
about the last combination). This is a very strong check, as at intermediate steps 
we get contributions from $c_{W_1}$, $c_{W_2}$, $c_{B_1}$, $d_6^{hl}$, $d_8^{hl}$, $d_{10}^{hl}$, $c_1^{hl}$ and $c_D$, which only 
at the end of the computation arrange themselves in gauge-independent combinations. 

The counterterm of each Wilson coefficient can be easily reconstructed from the RGEs knowing that the scaling with the renormalization scale 
is $\nu^{2\epsilon}$.

\section{Solution and numerical analysis}
\label{Sec:numerics}

We are only interested in the solution of the RGEs of gauge independent quantities, i.e. of those displayed in Sec. \ref{Sec:RGEPH}.
These RGEs can be rewritten in a compact form by defining a vector ${\bf A}=\{ \bar c_{B_1}, c_{B_2}, d_4^{hl}, d_5^{hl}, \bar d_8^{hl}, 
\bar d_{10}^{hl}, d_7^{hl}, d_9^{hl}, d_{11}^{hl}\}$ (we do not include the RGEs of $c_{p'p}$ and $\bar c_W$ because they are identical to the 
ones found in Ref. \cite{lmp}, and were already solved in the same reference. Indeed, their solution can be easily found using the reparametrization 
invariant relations found in Ref. \cite{Manohar:1997qy}. As pointed out in Sec. \ref{Sec:RGEPH}, we do not know if the Wilson 
coefficients $d_{7}^{hl}$, $d_{9}^{hl}$ and $d_{11}^{hl}$ are physical or not, but we will solve their RGEs anyway). They read 

\be
\label{rgecompact1}
\nu \frac{d {\bf A}}{d \nu} = \frac{\al}{\pi}({\bf M}  {\bf A}+{\bf F}(\al))
\,.
\ee
The matrix ${\bf M}$ and the vector ${\bf F}$ follow from the RGEs given in Sec. \ref{Sec:RGEPH}. The running of the strong coupling constant, $\al$, is 
needed only with LL accuracy:
\be
 \nu   {{\rm d \al} \over {\rm d} \nu}   
\equiv
\beta(\alpha_s)
=
-2 \al
\left\{\beta_0{\al \over 4 \pi} + \cdots\right\}
\,,
\ee
where 
\begin{eqnarray}
  \beta_0&=&{11 \over 3}C_A -{4 \over 3}T_Fn_f\,,
\end{eqnarray}
and $n_f$ is the number of dynamical (active) quarks i.e. the number of massless quarks. 

In this approximation, the Eq. (\ref{rgecompact1}) can be simplified to
\be
\frac{d {\bf A}}{d \al} = -\frac{2}{ \beta_0\al}({\bf M}  {\bf A} + {\bf F}(\al))
\,.
\ee
It is more convenient to define $z \equiv \left(\frac{\al(\nu)}{\al(m)}\right)^{\frac{1}{\beta_0}}\simeq
1-\frac{1}{2\pi}\al(\nu)\ln (\frac{\nu}{m})$ and rewrite the equation above as:

\begin{equation}
\label{definitiveRGE}
 \frac{d{\bf A}}{dz}=-\frac{2}{z}(\mathbf{M}{\bf A} + {\bf F}(z))\,.
\end{equation}
In order to solve Eq. \ref{definitiveRGE}, we need the initial matching conditions at the hard scale, at tree-level. For the bilinear sector, they 
have been determined in Ref.~\cite{Manohar:1997qy} and read $c_k=c_F=c_D=c_S=c_{W_1}=c_{B_1}=1$ and $c_{W_2}=c_{p'p}=c_{B_2}=0$. There are no tree level 
contribution to the 
Wilson coefficients associated to heavy-light operators, so its initial matching conditions are $c_i^{hl}=0,\,i=1,\ldots 4$ and 
$d_i^{hl}=0,\,i=4,\ldots,11$. The Wilson coefficients $c_k$, $c_F$, $c_S=2c_F -1$, $c_{p'p}=c_F-1$, $\bar c_W=1$, $\bar c_1^{hl}$, 
$c_2^{hl}$, $c_3^{hl}$ and $c_4^{hl}$ are needed with LL accuracy. They can be found in Refs. \cite{Pineda:2011dg,Manohar:1997qy,Bauer:1997gs}

After solving the RGEs we obtain the LL running of the Wilson coefficients associated to the $1/m^3$ spin-dependent operators 
of the HQET Lagrangian including spectator quark effects. The solution is numerical and reads

$$\bar c_{B_1} = -0.695 + \frac{0.045788}{z^{11.333333}} + \frac{13.6766}{z^{9.762121}}
 - \frac{10.4869}{z^{9}} + \frac{0.0690}{z^{8.333333}} 
 - \frac{0.3586}{z^{8.24892}} $$
\begin{equation}
 - \frac{1.83813}{z^{6.833333}}
 + \frac{1.34179}{z^{6.549986}}
 + \frac{1}{z^6} + \frac{4.49328}{z^{3.833333}}
 - \frac{4.95805}{z^{3.577865}} - \frac{3.290}{z^{3}}\,,
\end{equation}\\

$$ c_{B_2} = -2.966 - \frac{0.065821}{z^{11.333333}} - \frac{16.4471}{z^{9.762121}} 
+ \frac{13.4869}{z^9} - \frac{0.0493}{z^{8.333333}} 
+ \frac{0.2388}{z^{8.24892}}$$
\begin{equation}
 - \frac{2.94100}{z^{6.833333}} + \frac{4.6355}{z^{6.549986}} 
- \frac{4.3137}{z^6} - \frac{15.2080}{z^{3.833333}} 
+ \frac{16.0570}{z^{3.577865}} + \frac{7.572}{z^3}\,,
\end{equation}\\

$$ d_4^{hl} = 0.0407609 - \frac{3.203\cdot 10^{-23}}{z^{11.333333}} - \frac{3.9035\cdot 10^{-21}}{z^{9.5}}
- \frac{5.1587\cdot 10^{-22}}{z^9} + \frac{7.87\cdot 10^{-22}}{z^{8.333333}}$$
$$+ \frac{3.053\cdot10^{-21}}{z^{8.24892}} + \frac{2.2581\cdot 10^{-21}}{z^{6.833333}} 
+ \frac{3.8386\cdot 10^{-23}}{z^{6.549986}} + \frac{0.024038462}{z^{6}}$$
\begin{equation}
 - \frac{0.1897993}{z^{3.833333}} 
+ \frac{1.97644\cdot 10^{-20}}{z^{3.577865}} + \frac{0.125}{z^3}\,,
\end{equation}\\

$$ d_5^{hl} = 0.01 - \frac{2.22704\cdot 10^{-22}}{z^{11.333333}} - \frac{5.46119\cdot10^{-20}}{z^{9.762121}} 
+ \frac{1.24130\cdot 10^{-21}}{z^{9.5}}
+ \frac{4.30351\cdot 10^{-20}}{z^{9}} $$
$$- \frac{0.00851190}{z^{8.333333}}+ \frac{3.836\cdot10^{-23}}{z^{8.24892}} + \frac{1.10836\cdot10^{-21}}{z^{6.833333}}
- \frac{9.1837\cdot10^{-23}}{z^{6.549986}}  $$
\begin{equation}
- \frac{0.01190476}{z^{6}} 
+ \frac{5.1030\cdot 10^{-22}}{z^{3.833333}}
- \frac{1.84872\cdot10^{-21}}{z^{3.577865}}  + \frac{0.0104167}{z^{3}}\,,
\end{equation}\\

$$ \bar d_8^{hl} = -0.1149 + \frac{0.0069309}{z^{11.333333}} + \frac{0.206216}{z^{9.762121}} 
- \frac{0.211554}{z^{9}} - \frac{0.00663}{z^{8.333333} } + \frac{0.03645}{z^{8.24892}}$$
\begin{equation}
 + \frac{0.235510}{z^{6.833333}} - \frac{0.090907}{z^{6.549986}}
- \frac{0.24908}{z^6} - \frac{2.88931}{z^{3.833333}} 
+ \frac{3.27297}{z^{3.577865}} - \frac{0.1957}{z^3}\,,
\end{equation}\\

$$ \bar d_{10}^{hl} = -0.851 - \frac{0.000658}{z^{11.333333}} + \frac{1.52703}{z^{9.762121}} 
- \frac{1.3162}{z^{9.5}} - \frac{1.39575}{z^{9}} + \frac{0.2628}{z^{8.333333}}
+ \frac{1.024}{z^{8.24892}}$$
\begin{equation}
+ \frac{0.81541}{z^{6.833333}} + \frac{0.0125353}{z^{6.549986}} 
- \frac{0.1392}{z^6} - \frac{7.7003}{z^{3.833333}} 
+ \frac{8.65108}{z^{3.577865}} - \frac{1.890}{z^3}\,,
\end{equation}\\

$$ d_7^{hl} = -0.000867 + \frac{0.0000799}{z^{11.888889}} + \frac{3.346\cdot10^{-24}}{z^{11.333333}}
+ \frac{4.0769\cdot 10^{-22}}{z^{9.5}} + \frac{0.020957}{z^{9}}$$
$$- \frac{0.036600}{z^{8.333333}}- \frac{3.188\cdot10^{-22}}{z^{8.24892}} + \frac{0.01792}{z^{6.833333}}
- \frac{4.0091\cdot10^{-24}}{z^{6.549986}} $$
\begin{equation}
- \frac{0.000611}{z^{6}} - \frac{0.00086}{z^{3.833333}}
- \frac{2.06422\cdot 10^{-21}}{z^{3.577865}} - \frac{0.00002}{z^{3}}\,,
\end{equation}\\

$$d_9^{hl} = -0.006752 + \frac{0.0000479}{z^{11.888889}} - \frac{1.5783\cdot10^{-24}}{z^{11.333333}} 
- \frac{1.9232\cdot10^{-22}}{z^{9.5}} - \frac{0.009862}{z^9}$$
$$+ \frac{0.036600}{z^{8.333333}} 
+ \frac{1.504\cdot10^{-22}}{z^{8.24892}} 
- \frac{0.067979}{z^{6.833333}} 
+ \frac{1.89125\cdot10^{-24}}{z^{6.549986}} $$
\begin{equation}
+ \frac{0.019841}{z^{6}} + \frac{0.05322}{z^{3.833333}} 
+ \frac{9.7377\cdot10^{-22}}{z^{3.577865}} - \frac{0.025117}{z^3}\,,
\end{equation}\\

$$d_{11}^{hl} = -0.000769 - \frac{0.00001598}{z^{11.888889}} + \frac{7.019\cdot10^{-25}}{z^{11.333333}}
+ \frac{8.553\cdot10^{-23}}{z^{9.5}} + \frac{0.0006164}{z^9} $$
$$- \frac{0.006880}{z^{8.333333}}- \frac{6.69\cdot10^{-23}}{z^{8.24892}} + \frac{0.013287}{z^{6.833333}} 
- \frac{8.4104\cdot10^{-25}}{z^{6.549986}}$$
\begin{equation}
- \frac{0.009005}{z^6} + \frac{0.008295}{z^{3.833333}} 
- \frac{4.33038\cdot10^{-22}}{z^{3.577865}} - \frac{0.005529}{z^{3}}\,.
\end{equation}
The single log results can be found analytically by solving Eqs. (\ref{rged4hl})-(\ref{rged11hl}) just taking the tree level values of the 
Wilson coefficients that appear and considering $\alpha$ as a constant. We do not present the single log result of $\bar c_{B_1}$ and 
$c_{B_2}$ because spectators do not affect them, as their matching conditions are zero, and they were already found in Ref. \cite{lmp}. For the Wilson 
coefficients associated to heavy-light operators we obtain

\begin{equation}
d_{4}^{hl} = - \frac{1}{16} (8C_F-3C_A)\frac{\alpha}{\pi}\ln\left(\frac{\nu}{m}\right) 
+ \mathcal{O}(\alpha^2)\,,
\end{equation}

\begin{equation}
  d_{5}^{hl} = \frac{1}{8} C_F(2C_F-C_A)\frac{\alpha}{\pi}\ln\left(\frac{\nu}{m}\right) 
  + \mathcal{O}(\alpha^2)\,,
\end{equation}

\begin{equation}
 \bar d_{8}^{hl}(\nu) = \mathcal{O}(\alpha^2)\,,
\end{equation}

\begin{equation}
 \bar d_{10}^{hl} = -1 + \left(\frac{4}{3}C_F - \frac{5}{12}  C_A\right)\frac{\alpha}{\pi}\ln\left(\frac{\nu}{m}\right) 
  + \mathcal{O}(\alpha^2)\,,
\end{equation}

\begin{equation}
 d_{7}^{hl} = \mathcal{O}(\alpha^2)\,,
\end{equation}

\begin{equation}
  d_{9}^{hl} =  \mathcal{O}(\alpha^2)\,,
\end{equation}

\begin{equation}
  d_{11}^{hl} =  \mathcal{O}(\alpha^2)\,.
\end{equation}
Note that $\bar d_8^{hl}$ and $d_i^{hl},\,i=7,9,11$, are zero at the level of the single log. This means that 
the first contribution will be of $\mathcal{O}(\alpha^2\ln^2(\nu/m))$ and, as a consequence, their running will be small compared to the other 
Wilson coefficients because the single log dominates the expansion in the strong coupling, $\alpha$.

Spectator effects in HQET up to $\mathcal{O}(1/m^3)$ were already studied in Ref. \cite{Balzereit:1998am}. However, no anomalous dimension matrix for the 
Wilson coefficients was given, but only the single log expressions. At this level, we can compare our results with the ones given in that reference. The 
first thing we observe is that, in Ref. \cite{Balzereit:1998am}, it is stated that spin-dependent heavy-light operators change the single log results of the bilinear 
sector already found in Ref. \cite{Balzereit:1998jb}. That is strange, because the initial matching conditions of heavy-light operators is zero, and therefore, 
they should not change the single log expressions. After a more detailed comparison, taking the single logs given in Ref. \cite{Balzereit:1998am} 
and using Eqs.(47)-(51) of Ref. \cite{lmp} to change the operator basis, we find that, for physical combinations, the single log results remain 
unchanged and are still in agreement with Ref. \cite{lmp} and with what we find in this paper (that single logs remain unchanged including spectators). 
Concerning the running of heavy-light 
operators, we find that $c_{7+}^{(3h)o} = c_{7-}^{(3h)o}=8d_4^{hl}$. The first equality is already in disagreement with  Ref. \cite{Balzereit:1998am}, and 
for the explicit single logs given in it, only the term proportional to $C_F$ agrees with ours. We also find 
that $c_{7+}^{(3h)s}=c_{7-}^{(3h)s}=8d_5^{hl}$, which leads to agreement between the single logs presented in Ref. \cite{Balzereit:1998am} and ours. Also 
the given results for $d_7^{hl}$, $d_9^{hl}$ and $d_{11}^{hl}$ are in agreement. We find that 
$c_{6-}^{(3l)o} + c_{7-}^{(3l)o} = 8\bar d_8^{hl}$, which also agrees. Finally, we find that 
$\bar d_{10}^{hl} = c_6^{(3l)o}-(c_{7+}^{(3h)o}-c_{7-}^{(3h)o})/2-c_{W_1}^{FG}$ 
(where $c_{W_1}^{FG}$ is the Wilson coefficient $c_{W_1}$ evaluated in the Feynman gauge, whose single log expression was 
found in Ref. \cite{Balzereit:1998jb}), for which we find disagreement 
(even though a change of sign in the single log of $c_6^{(3l)o}$ plus the condition $c_{7+}^{(3h)o} = c_{7-}^{(3h)o}$, expected to 
reproduce $d_4^{hl}$ correctly, would lead to agreement. This also would imply a change of sign in the single log expression of $c_{7-}^{(3l)o}$).

\begin{figure}[!htb]
	\includegraphics[width=0.43\textwidth]{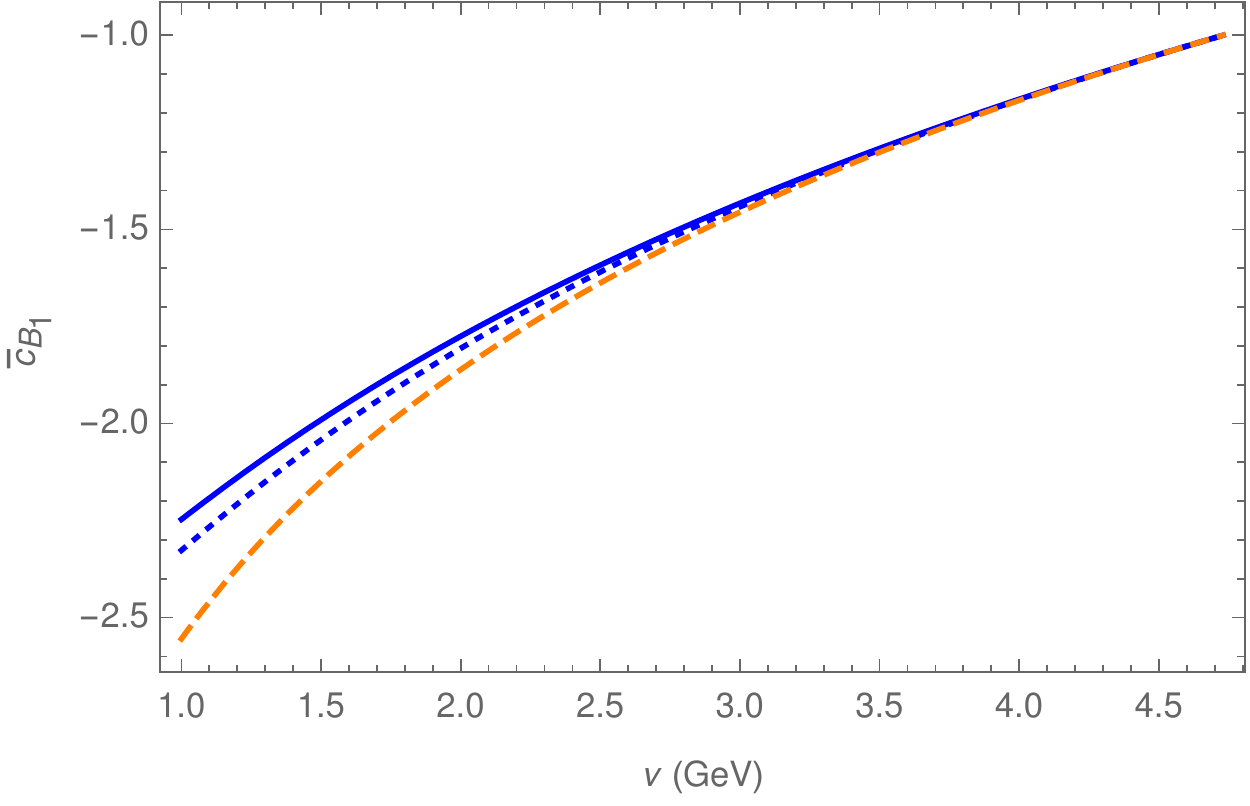}
\hspace{6ex}
	\includegraphics[width=0.43\textwidth]{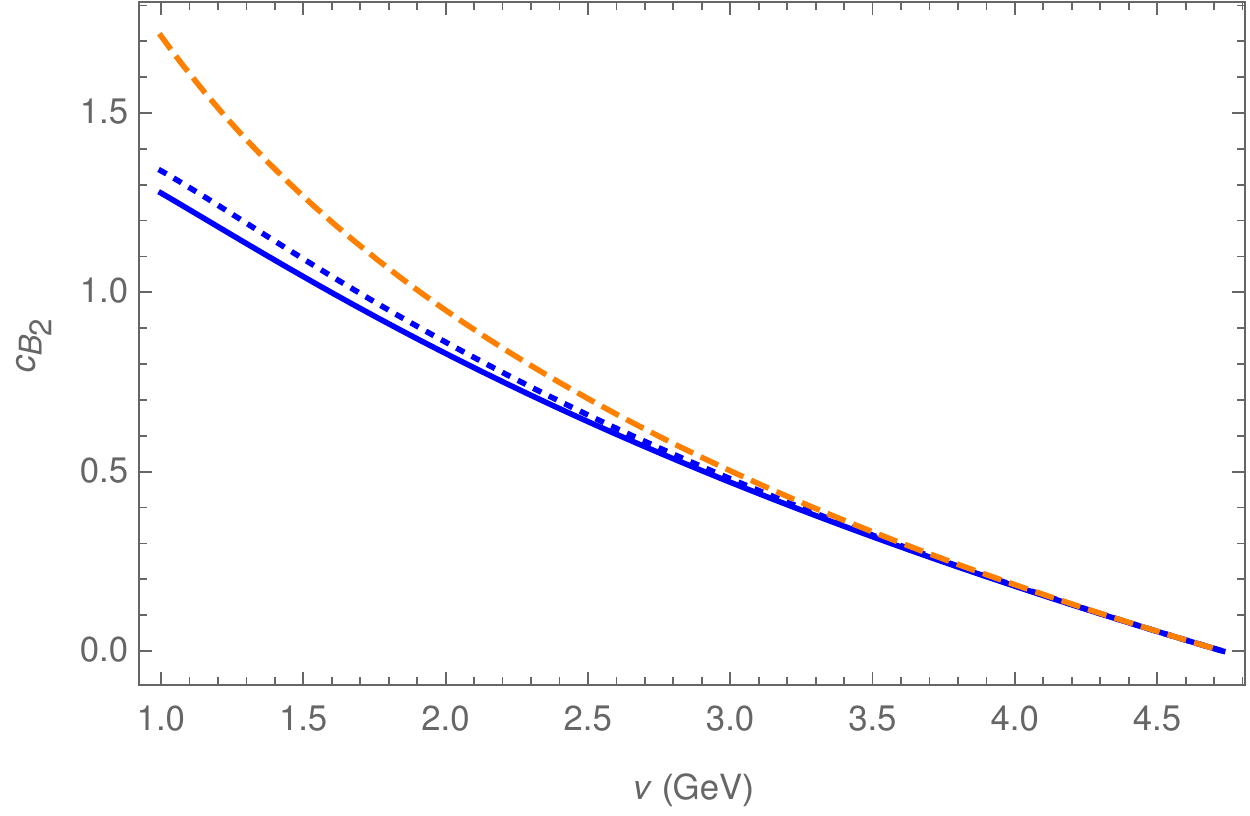}
	%
	\vspace{1ex}
	\includegraphics[width=0.43\textwidth]{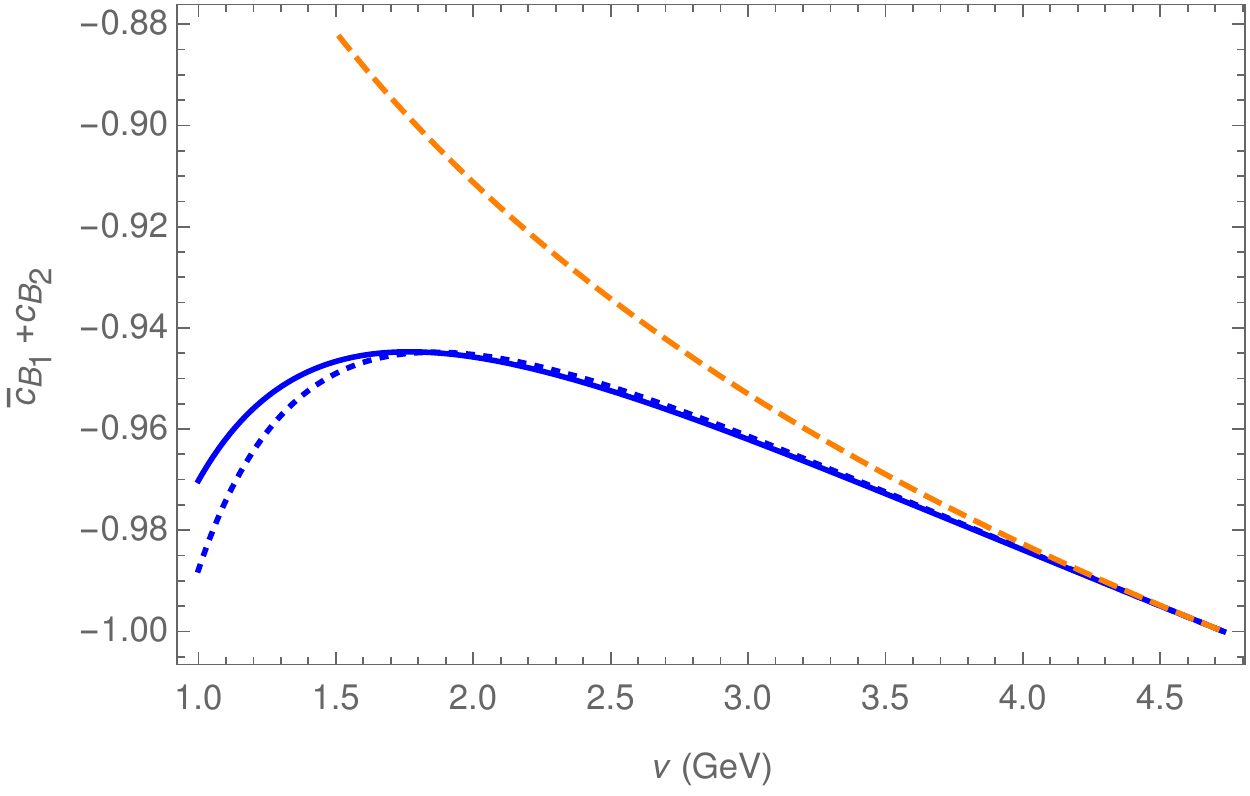}
\hspace{6ex}
	\includegraphics[width=0.43\textwidth]{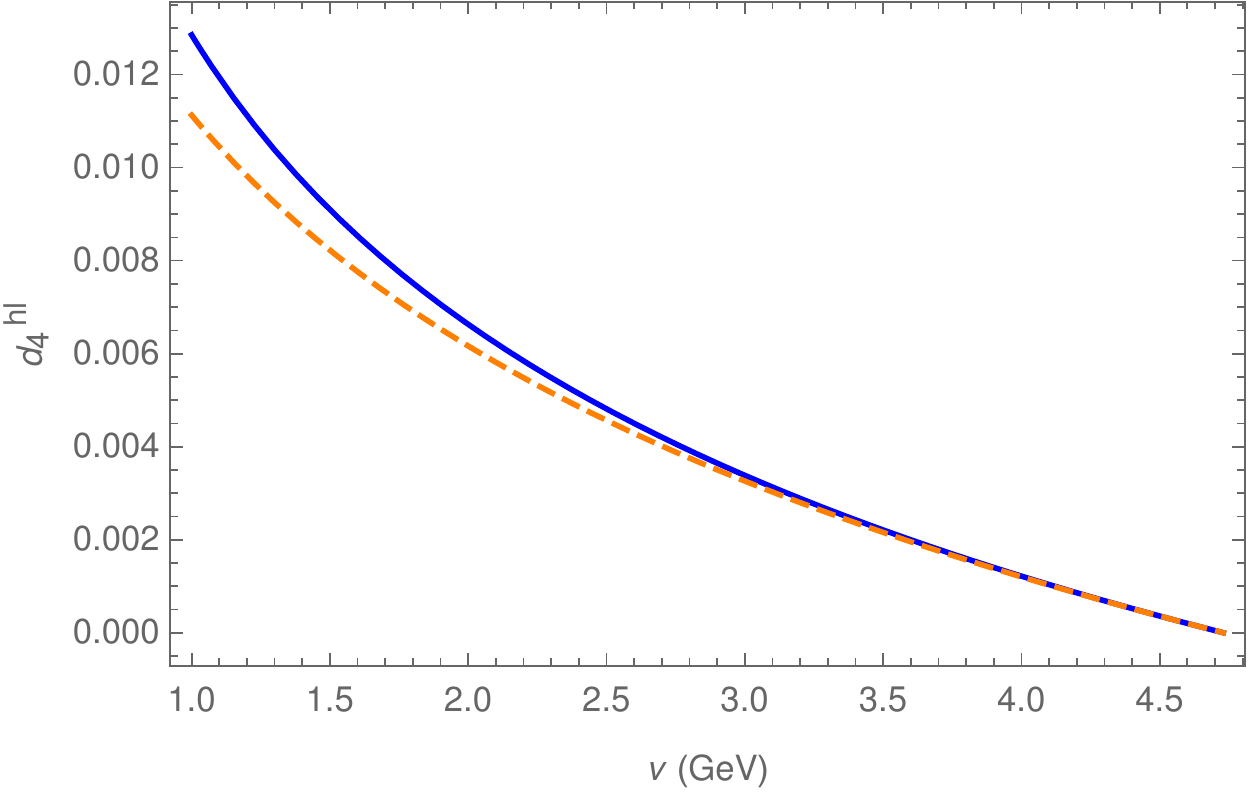}
	%
	\vspace{1ex}
	\includegraphics[width=0.43\textwidth]{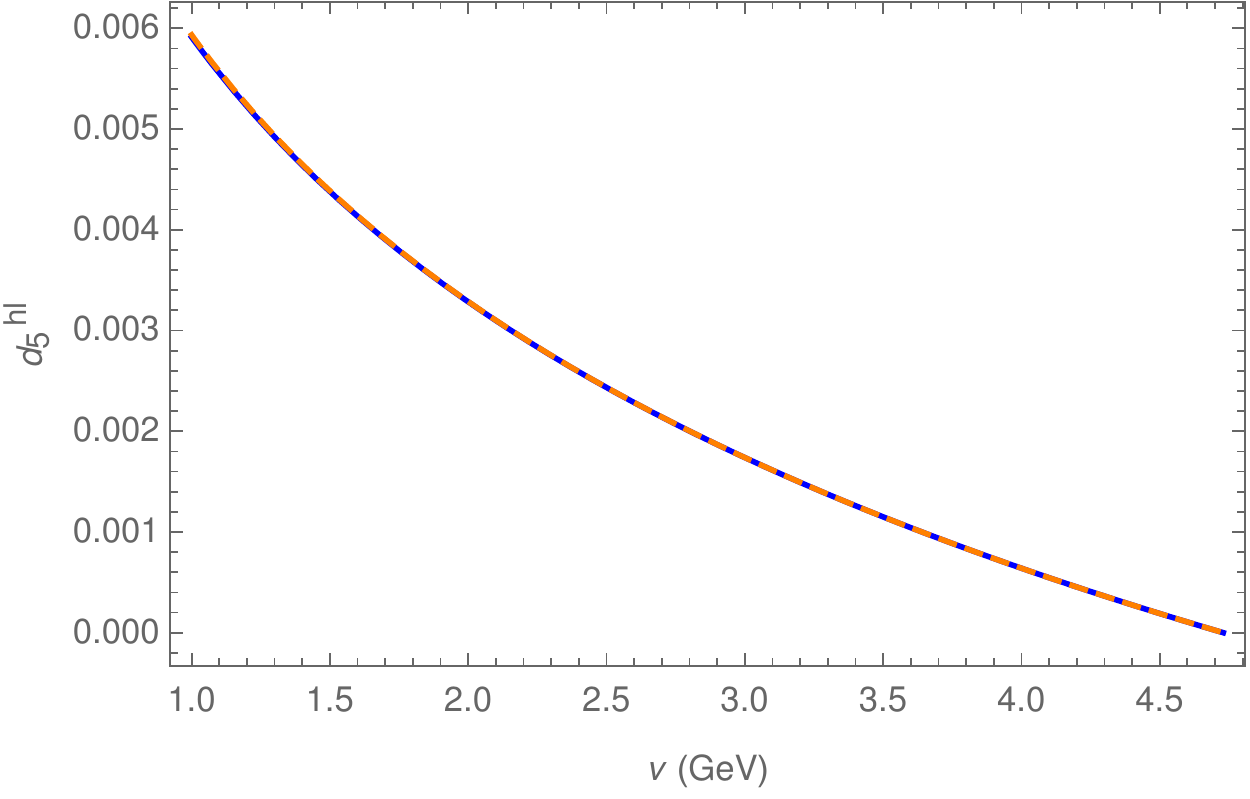}
\hspace{6ex}
	\includegraphics[width=0.43\textwidth]{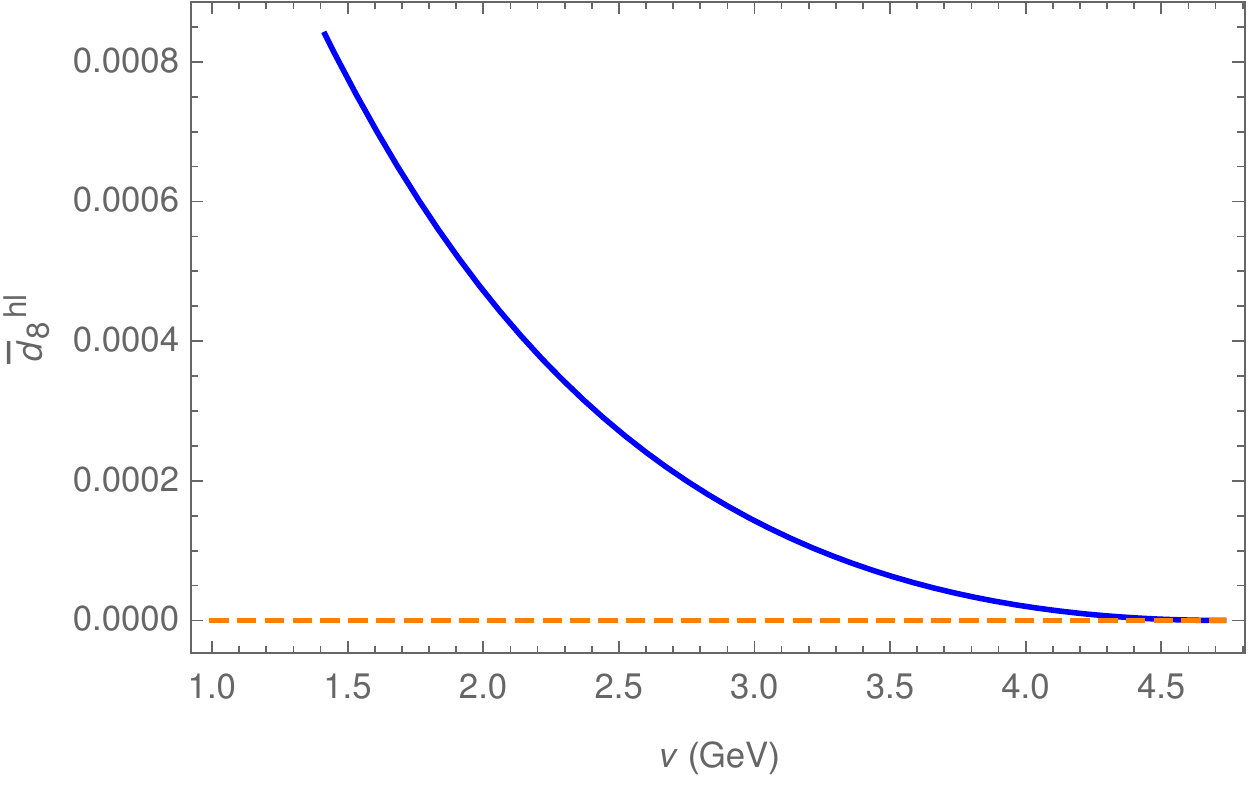}
\caption{Running of the $1/m^3$ spin-dependent Wilson coefficients: $\bar c_{B_1}$, $c_{B_2}$, $\bar c_{B_1} + c_{B_2}$, $d_4^{hl}$, $d_5^{hl}$ and 
$\bar d_8^{hl}$, applied to the bottom heavy quark case. The continuous line is the LL result with $n_f=4$, 
the dotted line is the LL result with $n_f=0$ and the dashed line is the single leading logarithmic result (it does not depend on $n_f$). 
\label{Plots1}}   
\end{figure}

\begin{figure}[!htb]
	\includegraphics[width=0.43\textwidth]{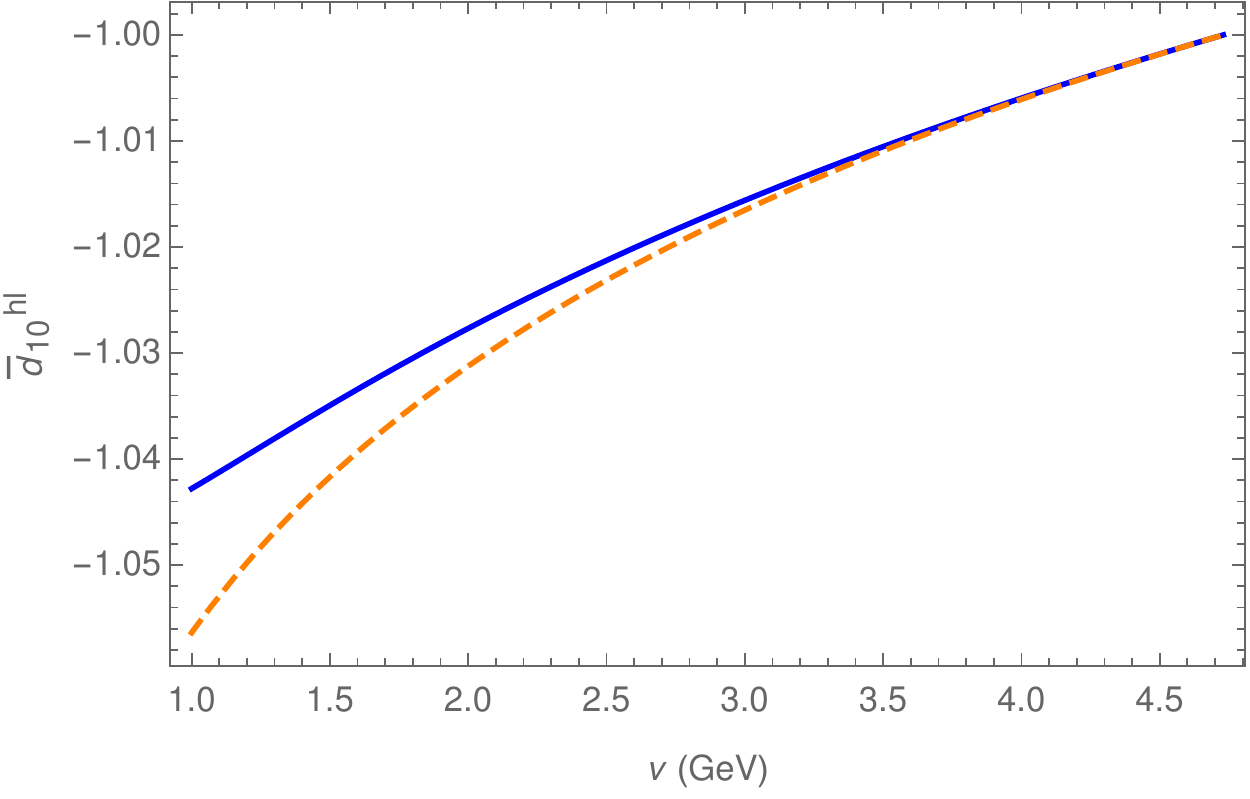}
\hspace{6ex}
	\includegraphics[width=0.43\textwidth]{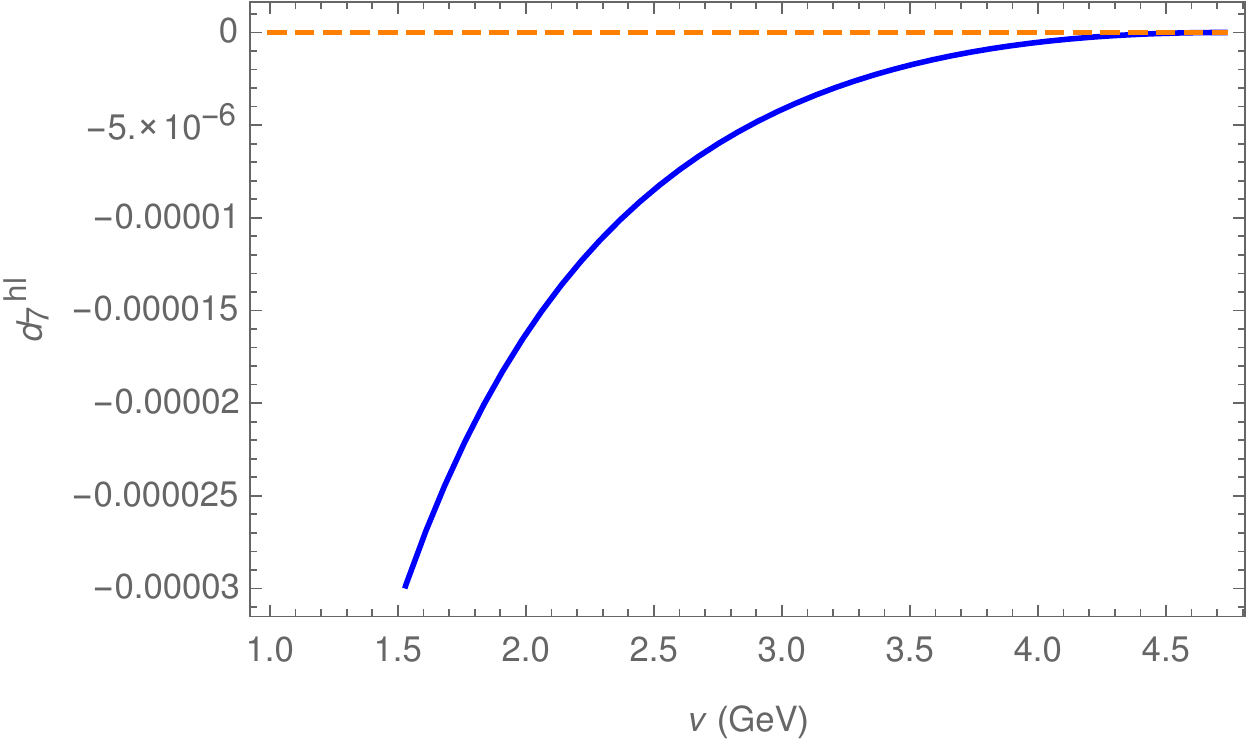}
	\vspace{1ex}
	\includegraphics[width=0.43\textwidth]{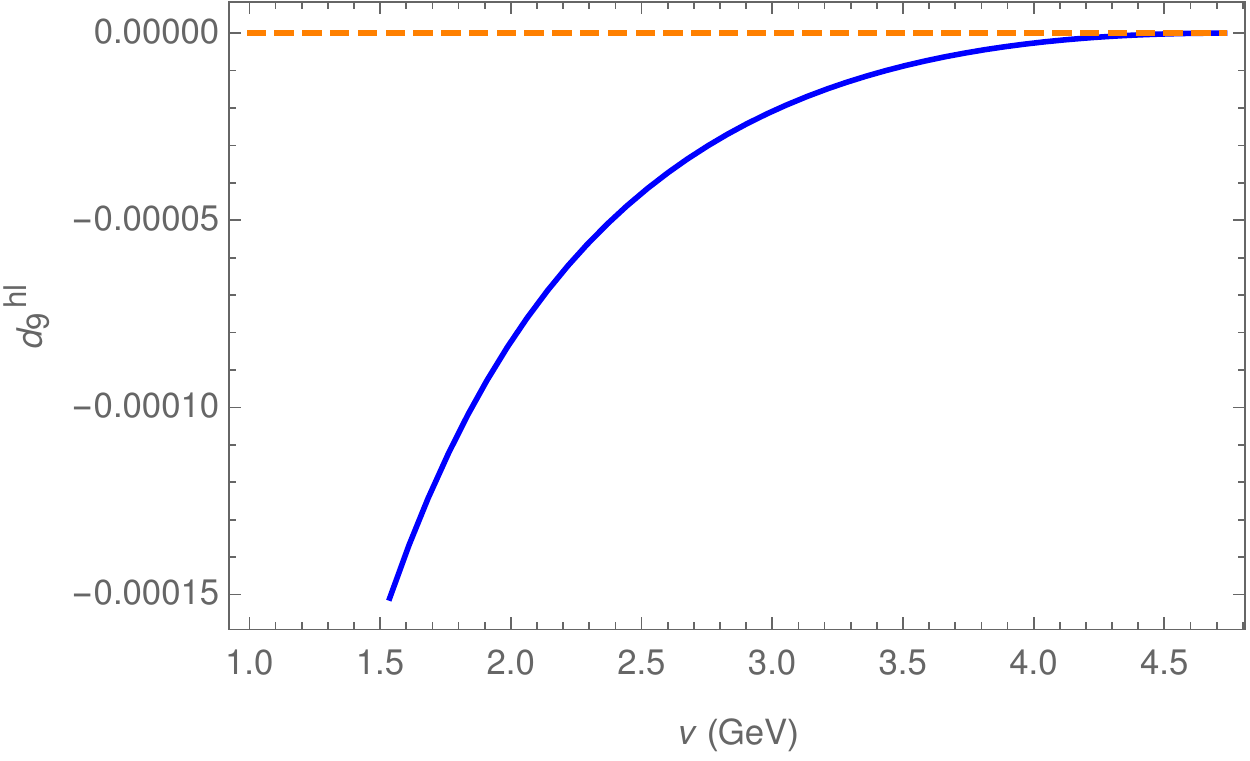}
\hspace{6ex}
	\includegraphics[width=0.43\textwidth]{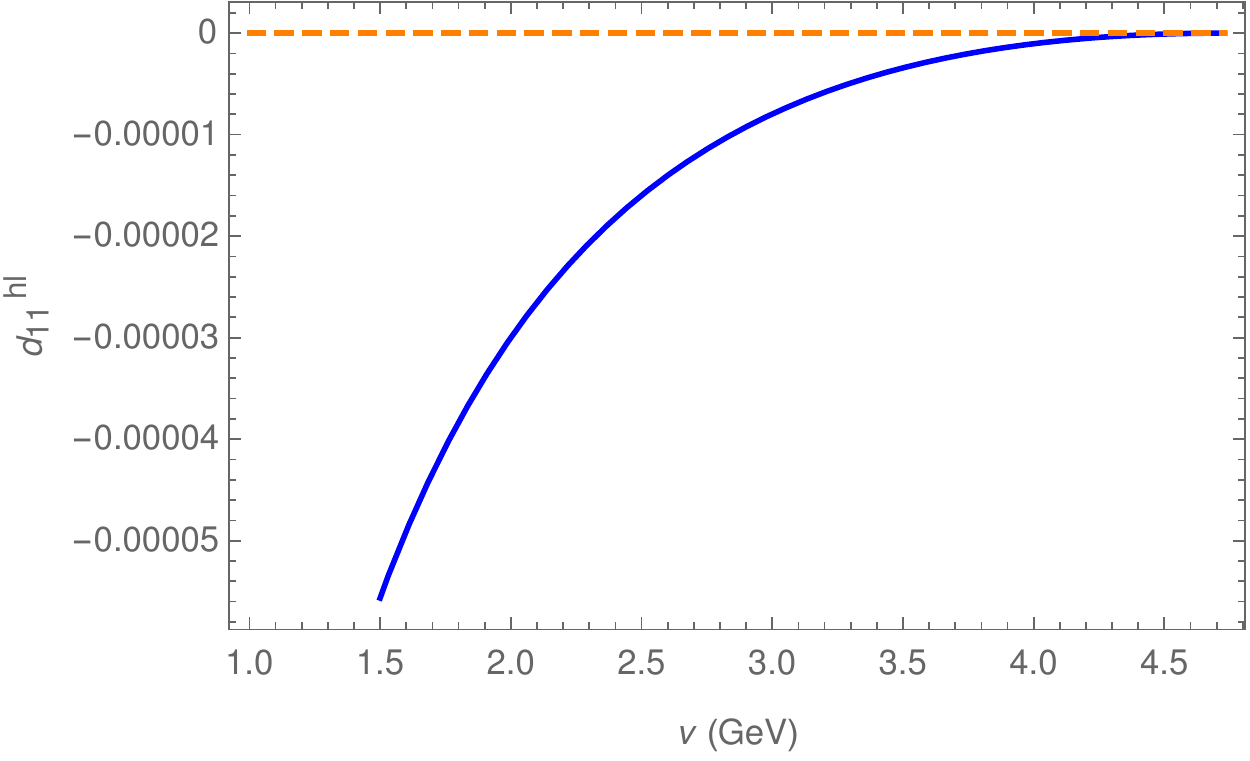}
\caption{Running of the $1/m^3$ spin-dependent Wilson coefficients: $\bar d_{10}^{hl}$, $d_7^{hl}$, $d_9^{hl}$ and 
$d_{11}^{hl}$, applied to the bottom heavy quark case. The continuous line is the LL result with $n_f=4$ and the dashed line is the 
single leading logarithmic result (it does not depend on $n_f$). 
\label{Plots2}}   
\end{figure}

In Figs. \ref{Plots1}, \ref{Plots2} we plot the results ontained in Sec. \ref{Sec:numerics} applied to the bottom heavy quark case, ilustrating 
the importance of incorporating large logarithms in heavy quark physics. Only physical combinations and specific combinations that appear 
in physical observables, like Compton scattering (see Ref. \cite{lmp}), are represented. We run the Wilson coefficients from the 
heavy quark mass to 1 GeV. For illustrative purposes, we take $m_b=4.73$ GeV and $\al(m_b)=0.215943$.

Concerning the numerical analysis, we observe that spector quarks change slighly the running of the physical quantities computed in Ref. \cite{lmp}, 
$\bar c_{B_1}$ and $c_{B_2}$, but that change is small (of approximately $0.1$ after running, with respect to the LL result with $n_f=0$, when they have
a value of $\sim -2$ and $\sim 1$, respectively), so the effect induced by them is numerically subleading. However, the effect induced by the
spectators tends to get away the curve from the single log one, so it makes the resummation of large logs more important. The change in combinations that appear in Compton scattering, like 
$\bar c_{B_1} + c_{B_2}$ is sizable, but even smaller than before. It changes by $0.02$ after running with respect to the LL result with $n_f=0$. Concerning 
the Wilson coefficients associated to heavy-light operators, we find that their running is small but sizable in some cases. The running is saturated 
by the single log in $d_4^{hl}$, $d_5^{hl}$ and $\bar d_{10}^{hl}$. In particular, $d_4^{hl}$ changes from $0$ to $0.012$ after 
running, and differs from the single log by $0.001$, $d_5^{hl}$ changes from $0$ to $0.006$. In that case, the resumation of logs happens to be 
unimportant. In the case of 
$\bar d_{10}^{hl}$, the resummation of logs introduces a difference of $\sim 0.015$ at $1$ GeV, with respect to the single log value. The 
Wilson coefficient runs from $-1$ to $-1.042$. The resumation of logs happens to be qualitatively very important for 
$\bar d_8^{hl}$, $d_{7}^{hl}$, $d_9^{hl}$ and $d_{11}^{hl}$, even though their running is small, because their behaviour is not saturated by the 
single log. They go from $0$ to $8.2\cdot 10^{-4}$, $-3\cdot 10^{-5}$, $-1.5\cdot 10^{-4}$ and $-5\cdot 10^{-5}$, respectively, after running 
at $\sim 1.5$ GeV.

\section{Conclusions}
\label{sec:conclusions}

We have obtained, for the first time, the LL running of the Wilson coefficients associated to the $1/m^3$ spin-dependent heavy-light operators of the HQET
Lagrangian, and their mixing with the Wilson coefficients associated to the $1/m^3$ spin-dependent operators bilinear in the heavy quark 
fields. It has been observed that, spectator quark effects are numerically subleading with respect to the ones coming from the bilinear sector. It has been 
proven that, after 
the inclusion of massless fermions, the relations coming from reparametrization invariance \cite{Manohar:1997qy} are still satisfied and that the 
running of physical quantities depends only on gauge-independent quantities, as expected. Even though spectator effects are found to 
be numerically subleading, they have to be included, formally.

The presented results are written in a more standard basis, set by Ref. \cite{Manohar:1997qy}, than the one used previously by 
Refs. \cite{Balzereit:1998am,Balzereit:1998jb,Balzereit:1998vh}, 
and they are connected more closely to observables, as the quantities computed here are gauge independent. We have compared our results with the 
previous work done in Refs. \cite{Balzereit:1998jb,Balzereit:1998am}. For the gauge invariant combinations we have 
computed in our paper, the single logs presented in these references are in agreement with ours, except for $d_4^{hl}$ and $\bar d_{10}^{hl}$. 

The Wilson coefficients computed in this paper could have many applications in heavy quark and heavy quarkonium physics. In particular, 
they are necessary ingredients to obtain the pNRQCD Lagrangian with next-to-next-to-next-to-next-to-leading order (NNNNLO) and 
with next-to-next-to-next-to-next-to-leading logarithmic (NNNNLL) accuracy, which is the necessary precision to determine the $\mathcal{O}(m\alpha^6)$ 
and the $\mathcal{O}(m\alpha^7\ln\alpha + m\alpha^8\ln^2\alpha + \ldots)$ heavy quarkonium spectrum. They also have applications 
in QED bound states like in muonic hydrogen. 

\medskip

{\bf Acknowledgments} \\
We thank Antonio Pineda for reading over the manuscript. 
This work was supported by the Spanish grants FPA2014-55613-P, FPA2017-86989-P and SEV-2016-0588.

\appendix

\section{HQET Feynman rules}
\label{sec:fr}

\begin{figure}[!htb]
	\includegraphics[width=1.0\textwidth]{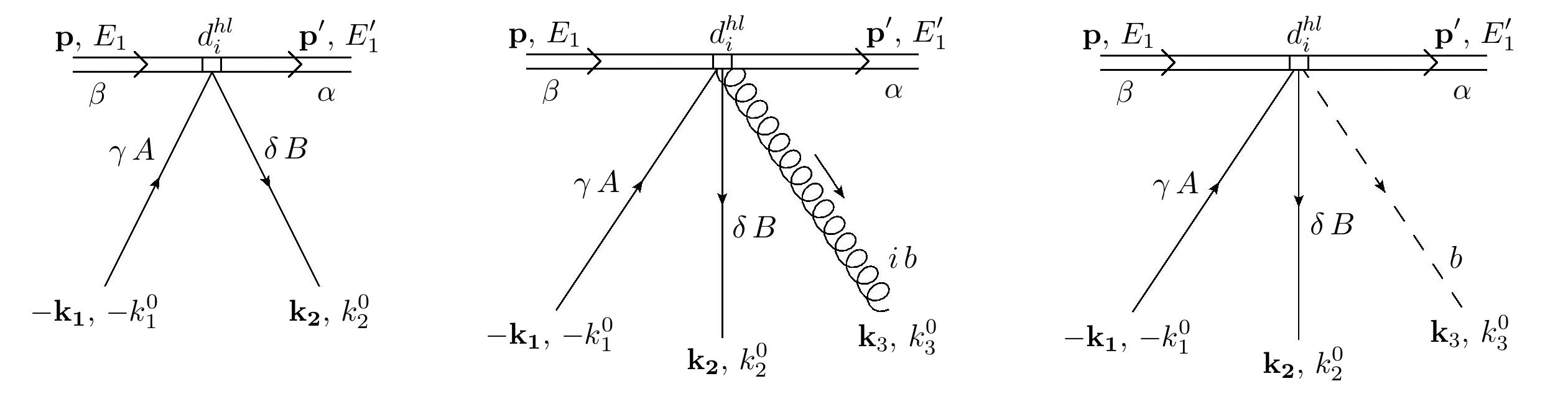}
\caption{Conventions for Feynman rules involving spin-dependent heavy-light operators which get LL runing. The double 
line represents a heavy quark, the single line a massless quark, and the curly and dashed lines represent a tranverse and longitudinal gluon respectively. 
The index $i$ goes from 4 to 11.}
\label{FeynmanRules}   
\end{figure}

Here we collect the new and necessary Feynman rules associated to the $1/m^3$ spin-dependent heavy-light operators given in Eqs. (\ref{O4})-(\ref{O11}), and 
that complement those that can be found in Refs. \cite{Pineda:2011dg,chromopolarizabilities}. The conventions are shown in Fig. \ref{FeynmanRules}.

\subsection{Proportional to $d_4^{hl}$}

\begin{equation}
 \mathcal{V} = - d_4^{hl}\frac{ig^2}{m^3}(\gamma^0 \gamma_5)_{BA} (T^a)_{\alpha\beta} (T^a)_{\delta\gamma} \bfsigma\cdot({\bf p} + {\bf p}')
\end{equation}

\begin{equation}
 \mathcal{V}^{i\,b} = d_4^{hl}\frac{ig^3}{m^3}(\gamma^0 \gamma_5)_{BA} (T^a)_{\delta\gamma} \{T^a,T^b\}_{\alpha\beta}\bfsigma^i
\end{equation}

\subsection{Proportional to $d_5^{hl}$}

\begin{equation}
 \mathcal{V} = - d_5^{hl}\frac{ig^2}{m^3}(\gamma^0 \gamma_5)_{BA} (I_{N_c})_{\alpha\beta} (I_{N_c})_{\delta\gamma}\bfsigma\cdot({\bf p}+{\bf p}')
\end{equation}

\begin{equation}
 \mathcal{V}^{i\,b} = d_5^{hl}\frac{2ig^3}{m^3}(\gamma^0 \gamma_5)_{BA} (I_{N_c})_{\delta\gamma} (T^b)_{\alpha\beta}\bfsigma^i
\end{equation}

\subsection{Proportional to $d_6^{hl}$}

\begin{equation}
 \mathcal{V} = d_6^{hl}\frac{ig^2}{m^3}(\gamma^0 \gamma_5)_{BA} (T^a)_{\alpha\beta} (T^a)_{\delta\gamma}\bfsigma\cdot({\bf k}_1-{\bf k}_2)
\end{equation}

\begin{equation}
 \mathcal{V}^{i\,b} = d_6^{hl}\frac{ig^3}{m^3}(\gamma^0 \gamma_5)_{BA} \{T^a,T^b\}_{\delta\gamma} (T^a)_{\alpha\beta}\bfsigma^i
\end{equation}

\subsection{Proportional to $d_{7}^{hl}$}

\begin{equation}
 \mathcal{V} = d_{7}^{hl}\frac{ig^2}{m^3}(\gamma^0 \gamma_5)_{BA} (I_{N_c})_{\alpha\beta} (I_{N_c})_{\delta\gamma}\bfsigma\cdot({\bf k}_1-{\bf k}_2)
\end{equation}

\begin{equation}
 \mathcal{V}^{i\,b} = d_{7}^{hl}\frac{2ig^3}{m^3}(\gamma^0 \gamma_5)_{BA} (T^b)_{\delta\gamma} (I_{N_c})_{\alpha\beta}\bfsigma^i
\end{equation}

\subsection{Proportional to $d_{8}^{hl}$}

\begin{equation}
 \mathcal{V} = d_{8}^{hl}\frac{ig^2}{m^3}(\gamma^i \gamma_5)_{BA} (T^a)_{\alpha\beta} (T^a)_{\delta\gamma}\bfsigma^i (k_1^0-k_2^0)
\end{equation}

\begin{equation}
 \mathcal{V}^{b} = d_{8}^{hl}\frac{ig^3}{m^3}(\gamma^i \gamma_5)_{BA} \{T^a,T^b\}_{\delta\gamma} (T^a)_{\alpha\beta}\bfsigma^i
\end{equation}

\subsection{Proportional to $d_{9}^{hl}$}

\begin{equation}
 \mathcal{V} = d_{9}^{hl}\frac{ig^2}{m^3}(\gamma^i \gamma_5)_{BA} (I_{N_c})_{\alpha\beta} (I_{N_c})_{\delta\gamma}\bfsigma^i (k_1^0-k_2^0)
\end{equation}

\begin{equation}
 \mathcal{V}^{b} = d_{9}^{hl}\frac{2ig^3}{m^3}(\gamma^i \gamma_5)_{BA} (T^b)_{\delta\gamma} (I_{N_c})_{\alpha\beta}\bfsigma^i
\end{equation}

\subsection{Proportional to $d_{10}^{hl}$}

\begin{equation}
 \mathcal{V} = d_{10}^{hl}\frac{g^2}{m^3} (T^a)_{\alpha\beta} (T^a)_{\delta\gamma}(\gamma^i)_{BA}(\bfsigma\times{\bf k})^i
\end{equation}

\begin{equation}
 \mathcal{V}^{i\,b} = d_{10}^{hl}\frac{g^3}{m^3}(\gamma^j)_{BA} [T^a,T^b]_{\delta\gamma} (T^a)_{\alpha\beta}\epsilon^{ijk}\bfsigma^k
\end{equation}

\subsection{Proportional to $d_{11}^{hl}$}

\begin{equation}
 \mathcal{V} = d_{11}^{hl}\frac{g^2}{m^3} (I_{N_c})_{\alpha\beta} (I_{N_c})_{\delta\gamma}(\gamma^i)_{BA}(\bfsigma\times{\bf k})^i
\end{equation}


\end{document}